\documentclass[prl,twocolumn,reprint,showpacs]{revtex4-1}
\pdfoutput=1
\usepackage{graphics,graphicx,amsfonts,amsmath,amsbsy,amssymb,color}
\usepackage[singlelinecheck=false]{subfig}
\usepackage{verbatim}  
\usepackage{psfrag}  
\usepackage{tabularx}
\usepackage{xmpmulti}
\usepackage[justification=justified]{caption}
\usepackage{hyperref}
\usepackage{pdfpages}

\newcommand{\refeq}[1]{{Eq.~(\ref{#1})}}
\newcommand{\reffig}[1]{{Fig.~\ref{#1}}}

\newcommand{\appropto}{\mathrel{\vcenter{
  \offinterlineskip\halign{\hfil$##$\cr
    \propto\cr\noalign{\kern2pt}\sim\cr\noalign{\kern-2pt}}}}}

\hypersetup{hyperindex,breaklinks,colorlinks=true,linkcolor=blue,citecolor=blue,urlcolor=magenta,filecolor=blue}
\allowdisplaybreaks

\begin{document}

\title{Range Separated Brueckner Coupled Cluster Doubles Theory}

\author{James~J.~Shepherd}
\email{jjs6@rice.edu}
\affiliation{Department of Chemistry and Department of Physics and Astronomy, Rice University, Houston, TX 77005-1892}
\author{Thomas~M.~Henderson}
\affiliation{Department of Chemistry and Department of Physics and Astronomy, Rice University, Houston, TX 77005-1892}
\author{Gustavo~E.~Scuseria}
\affiliation{Department of Chemistry and Department of Physics and Astronomy, Rice University, Houston, TX 77005-1892}


\begin{abstract}
We introduce a range-separation approximation to coupled cluster doubles (CCD) theory that successfully overcomes limitations of regular CCD when applied to the uniform electron gas. We combine the short-range ladder channel with the long-range ring channel in the presence of a Bruckner renormalized one-body interaction and obtain ground-state energies with an accuracy of 0.001 a.u./electron across a wide range of density regimes. Our scheme is particularly useful in the low-density and strongly-correlated regimes, where regular CCD has serious drawbacks. Moreover, we cure the infamous overcorrelation of approaches based on ring diagrams (i.e. the particle-hole random phase approximation). Our energies are further shown to have appropriate basis set and thermodynamic limit convergence, and overall this scheme promises energetic properties for realistic periodic and extended systems which existing methods do not possess.
\end{abstract}
\date{\today}
\maketitle

{\bf\emph{Introduction.--} }
The particle-hole random phase approximation (RPA) has a long history in condensed matter physics~\cite{bohm_collective_1951,*pines_collective_1952,*bohm_collective_1953}.  Recently, it has seen increased emphasis in the framework of density functional theory (DFT) for electronic structure, where its orbital-based description of electronic correlation has been of great utility~\cite{harl_accurate_2009,leininger_combining_1997, *goll_short-range_2005,*goll_short-range_2006,*lebegue_cohesive_2010}.   One of the most important successes of the RPA in this context lies in the description of dispersion interactions~\cite{dobson_asymptotics_2006}, often in range-separated schemes which combine a long-range RPA picture with short-range DFT~\cite{toulouse_adiabatic-connection_2009,janesko_long-range-corrected_2009,bruneval_range-separated_2012}.  These range-separated schemes take advantage of the fact that the RPA accurately captures long-range correlations, and tend to mitigate its basis set sensitivity and its known failures for short-range correlations.  This failure occurs because RPA predicts too deep a correlation hole and therefore too large a correlation energy~\cite{singwi_electron_1968,kurth_density-functional_1999}.
One popular remedy is the incorporation of second-order screened exchange (SOSEX) effects, as introduced by Freeman~\cite{freeman_coupled-cluster_1977}. This built on previous work by Gell-Mann and Brueckner, who demonstrated that second-order exchange is required to yield the correct constant term in the high-density expansion for the electron gas~\cite{gell-mann_correlation_1957}.  This correction has in recent times been adapted for a wide range of real electronic structure problems~\cite{harl_accurate_2009,eshuis_electron_2012,paier_assessment_2012}.

An alternative approach is motivated by noting that the RPA is an infinite-order summation of the ring diagrams from many-body perturbation theory~\cite{gell-mann_correlation_1957}.  From this perspective, the overbinding of the RPA is due to the absence of other diagrams.  Amongst these, the ladder diagrams are thought to be the most important~\cite{bishop_electron_1978,*bishop_electron_1982}.  Ladder diagrams have been proposed to yield high-quality short-range correlation functions and have been discussed as an alternative to screened exchange~\cite{bishop_electron_1978,*bishop_electron_1982,*bishop_overview_1991}.  However, ladder-only theories suffer from the divergences associated with finite-order perturbation theory applied to a Coulombic interaction in metallic solids%
~\cite{freeman_coupled-cluster_1983,cioslowski_applicability_2005,drummond_quantum_2009}.

We here suggest combining the long-range accuracy of RPA with the short-range accuracy of ladder diagrams within a range-separated CCD, based on modern developments relating RPA to CCD~\cite{scuseria_ground_2008,peng_equivalence_2013,scuseria_particle-particle_2013,van_aggelen_exchange-correlation_2013}. We perform a range separation with a Yukawa potential and apply this technique to the electron gas, a model with substantial historical significance~\cite{ceperley_ground_1980,perdew_self-interaction_1981} but nonetheless still receiving prominent attention~\cite{huotari_momentum_2010,drummond_quantum_2011,holzmann_momentum_2011,baguet_hartree-fock_2013}.  We show that this method improves both the undercorrelation of CCD and the overcorrelation of RPA, and does so in a manner which accurately captures the high momentum (short distance) approach to the complete basis set limit. We showcase high quality ($10^{-3}$~a.u./electron) results at a wide range of densities for finite electron numbers. We also demonstrate this method is suitable for studying the thermodynamic limit.%

{\bf\emph{Theory.--} } Coupled-cluster doubles (CCD) writes the correlated wave function $|\Psi\rangle$ in an exponential form:
\begin{equation}
|\Psi\rangle = \mathrm{e}^T |0\rangle \quad ; \quad T = \frac{1}{4} \, t_{ij}^{ab} \, a^\dagger_a \, a^\dagger_b \, a_j \, a_i,
\end{equation}
where $|0\rangle$ is a mean-field reference state (typically chosen to be Hartree--Fock), $T$ is an excitation operator and $t$ are amplitudes to be found. Here and throughout, indices $i$, $j$, $k$, $l$ label single-particle states occupied in $|0\rangle$ and $a$, $b$, $c$, $d$ label states unoccupied in $|0\rangle$; summation convention is used on dummy indices. 
Inserting this ansatz into the Schr\"odinger equation leads to an energy
\begin{align}
E &= \langle 0 | H | 0\rangle + E_\mathrm{corr} \quad ; \quad
E_\mathrm{corr} &= \frac{1}{4} \, t_{ij}^{ab} \, \bar{v}^{ij}_{ab},
\end{align}
and amplitude equation
\begin{align}
(\epsilon_i +\epsilon_j &-\epsilon_a -\epsilon_b) t_{ij}^{ab}=\bar{v}_{ij}^{ab} \label{ampeq} \\
&+\bar{v}_{cj}^{kb} t_{ik}^{ac} + \bar{v}_{ci}^{ka} t_{jk}^{bc} + \bar{v}_{cd}^{kl} t_{lj}^{db} t_{ik}^{ac} \notag \\
&+\frac{1}{2} \bar{v}_{cd}^{ab} t_{ij}^{cd} +\frac{1}{2} \bar{v}_{ij}^{kl}  t_{kl}^{ab} + \frac{1}{4} \bar{v}_{cd}^{kl} t_{ij}^{cd} t_{kl}^{ab} \notag \\
&- \bar{v}_{cj}^{ka} t_{ik}^{bc} - \bar{v}_{ci}^{kb} t_{jk}^{ac} - \bar{v}_{cd}^{kl} t_{lj}^{da} t_{ik}^{bc} \notag \\
&+\frac{1}{2} \bar{v}_{cd}^{kl} \left[  t_{lj}^{ab} t_{ik}^{cd} - t_{li}^{ab} t_{jk}^{cd} + t_{ji}^{db} t_{kl}^{ac} -t_{ij}^{da} t_{kl}^{bc} \right] \notag,
\end{align}
where $\epsilon$ are Hartree-Fock single-particle energies and $\bar{v}_{ij}^{ab} = v_{ij}^{ab} - v_{ij}^{ba}$ is the matrix representation of the electron-electron interaction. Further details can be found in a general review of coupled cluster theory~\cite{bartlett_coupled-cluster_2007}.%

The terms on the first line of the amplitude equation, \refeq{ampeq} are known as driving terms and arise in a lead-order perturbative expansion of the wave function.  
The terms on the second line are ring terms which together with the driving terms yield the particle-hole (ph) RPA.  Loosely, they treat electronic excitations as bosonic harmonic oscillators. 
The terms on the third line are ladder terms which together with the driving terms yield the particle-particle (pp) RPA, 
where they treat pairing excitations as a bosonic oscillator. We note that pp-RPA is probably a lesser-known type of RPA nonetheless familiar to physicists in nuclear structure theory~\cite{ring_nuclear_1980}.%

We call the terms on the fourth line ``crossed-ring" terms as they are ring-like but are needed to restore the fermionic antisymmetry of the $T$-amplitudes. The the last class of diagrams are what we name the ``mosaic" terms which are
joint ladder-ring diagrams that contribute to the Brueckner effective one-body Hamiltonian~\cite{scuseria_connections_1995}. 
Explicitly, we can incorporate these terms into the left hand side of the \refeq{ampeq} as follows:
\begin{align}
\label{bh1} \eta_i = \epsilon_i + \frac{1}{2} \bar{v}_{cd}^{il}t_{il}^{cd} \quad ; \quad \eta_a = \epsilon_a - \frac{1}{2} \bar{v}_{ad}^{kl}  t_{kl}^{ad}  .
\end{align}
Here, we have anticipated this matrix being diagonal in a canonical plane wave basis set due to momentum conservation. The interpretation of this is a renormalization of the single-particle spectrum to account for the effects of correlation. Overall, therefore, coupled-cluster theory combines these channels through a unified set of amplitudes and yields a properly fermionic wave function which incorporates both kinds of RPA fluctuations~\cite{scuseria_particle-particle_2013}. 

Because the pp-RPA accurately describes short-range correlations but underestimates long-range correlations, while the conventional ph-RPA is instead accurate for long-range correlations and overestimates short-range correlation, a natural approach which combines the benefits of both is to work through a range-separated scheme in which the Coulomb potential is split into a short-range piece to be treated with the ladder diagrams and a long-range piece which is treated with the ring terms.  We choose to use a Yukawa separation
\begin{equation}
\frac{1}{r_{12}} = \frac{\mathrm{e}^{-\gamma \, r_{12}}}{r_{12}} + \frac{1 - \mathrm{e}^{-\gamma \, r_{12}}}{r_{12}}
\end{equation}
where the parameter $\gamma$ is taken to be the Thomas-Fermi screening parameter. This represents the static limit of the screening from particle-hole RPA and is physically appropriate for metals, though it is well know that this approach is insufficiently screened in semiconductors or insulators~\cite{shimazaki_band_2008,leininger_combining_1997,*goll_short-range_2005,*goll_short-range_2006}.

We can now re-write the CCD amplitude equations replacing the potential with a long-range potential for the ring terms and a short-range potential for the ladder terms and excluding the crossed-rings:
\begin{equation}
\begin{split}
(\eta_i &+\eta_j-\eta_a -\eta_b) t_{ij}^{ab}=\bar{v}_{ij}^{ab} \\
&+\frac{1}{2} ( \bar{v}_\textrm{SR} ) {}_{cd}^{ab} t_{ij}^{cd} +\frac{1}{2} ( \bar{v}_\textrm{SR} ) {}_{ij}^{kl}  t_{kl}^{ab} + \frac{1}{4} ( \bar{v}_\textrm{SR} ) {}_{cd}^{kl} t_{ij}^{cd} t_{kl}^{ab} \\
&+( \bar{v}_\textrm{LR} ) {}_{cj}^{kb} t_{ik}^{ac} + ( \bar{v}_\textrm{LR} ) {}_{ci}^{ka} t_{jk}^{bc} + ( \bar{v}_\textrm{LR} ) {}_{cd}^{kl} t_{lj}^{db} t_{ik}^{ac} ,
\end{split}
\end{equation}
From the perspective of the random phase approximation, this is the addition of particle-particle terms with particle-hole terms~\cite{scuseria_particle-particle_2013}. 
The crossed-rings are not present in these equations because they do not represent an harmonic oscillator problem that can be added into pp-RPA or ph-RPA~\cite{scuseria_particle-particle_2013}.
Further technical details are provided as supplementary information~\footnote{See supplementary information.}.

\begin{figure}
\includegraphics[width=0.46\textwidth]{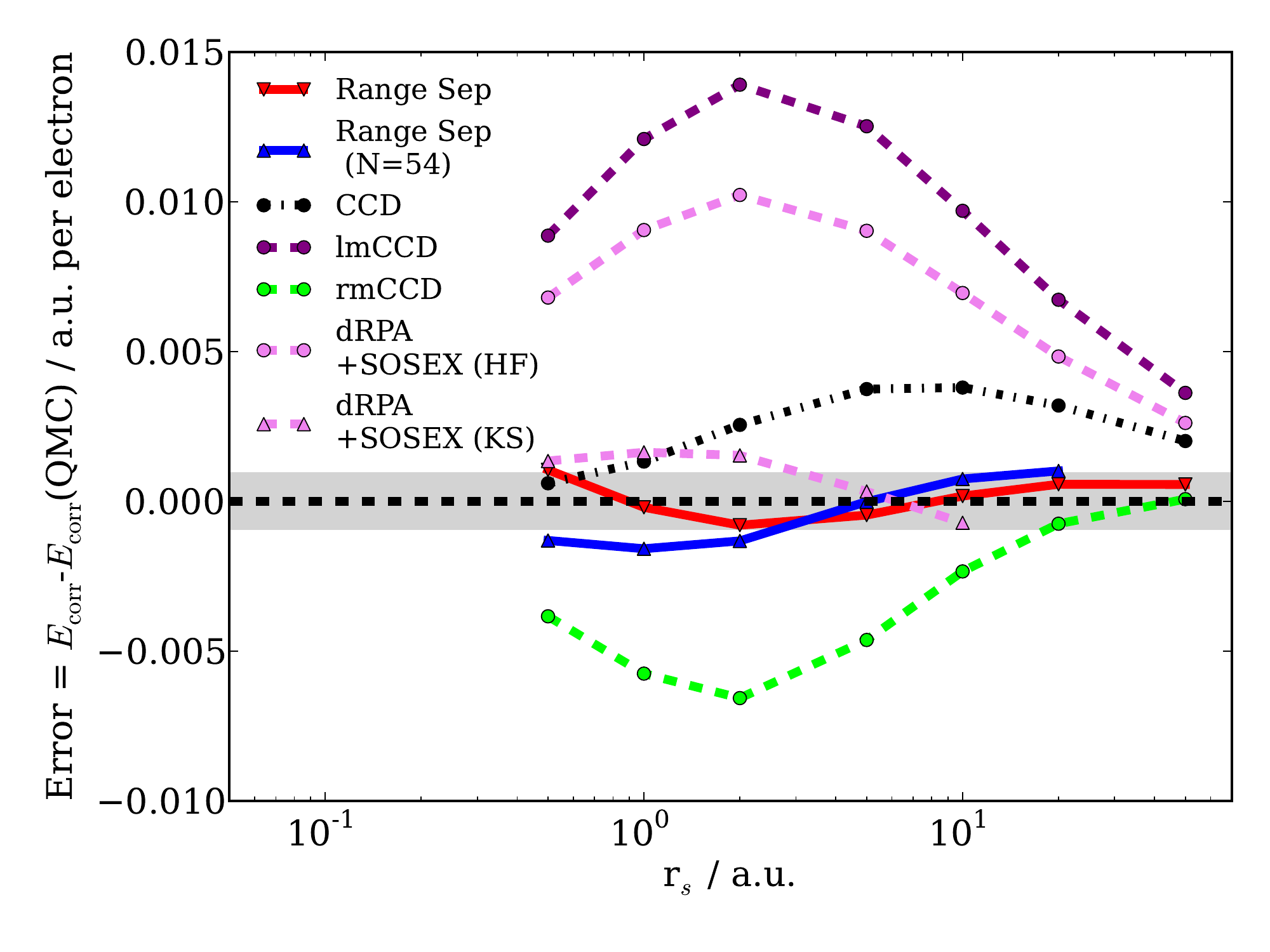}
\caption{Error of the range separated schemes measured with respect to QMC data; the shaded section of the graph represents 0.001 a.u. per electron. (Unless otherwise noted, $N$=14 and $M\rightarrow\infty$).}
\label{fig1}
\end{figure}

{\bf{\emph{Finite basis set electron gas calculations}.}--}  The continuum electron gas Hamiltonian is familiar to most, but here it is important to note that we use a basis of plane waves: $\phi_j \propto \exp({i {\bf k}_j \cdot {\bf r}}$).

The momenta, ${\bf k}$, associated with these plane waves have a finite spacing specified uniquely by a particle number ($N$) and a density ($r_s$, the Wigner-Seitz radius). 
This model is then typically referred to as the \emph{finite} simulation-cell electron gas. 
The infinite particle, $N\rightarrow\infty$, limit is referred to as the thermodynamic limit and the energy in this limit is uniquely defined only by the density $r_s$. 
In a frequently used procedure, diffusion Monte Carlo simulations can be made of finite electron numbers and extrapolated to the thermodynamic limit~\cite{ceperley_ground_1980,drummond_finite-size_2008}.

For the methods considered here, we also require an energy cutoff to give a finite number of basis functions ($M$); even for a finite particle number an infinite basis is required. 
The extrapolation procedure to reach the complete basis set limit, $M\rightarrow\infty$, has only been recently codified for plane waves~\cite{shepherd_convergence_2012,gruneis_explicitly_2013} although these methods have long been in use in quantum chemistry~\cite{hattig_explicitly_2012}.

In summary, the energies we are able to compute have three parameters: $M$, $N$ and $r_s$. 

{\bf\emph{Finite particle, complete basis set limit.--} }
We present results for the 14 electron system at the complete basis set limit in \reffig{fig1}, where we calculate the error of each method shown with respect to quantum Monte Carlo benchmarks~\footnote{For $N=14$, the benchmarks here come from full configuration interaction quantum Monte Carlo~\cite{booth_towards_2013} for the high-to-metallic density regime, and diffusion Monte Carlo for the remainder~\cite{needs_continuum_2010,lopez_rios__2013}. These results are effectively exact following extensive development and benchmarking~\cite{shepherd_full_2012,shepherd_investigation_2012,shepherd_convergence_2012,shepherd_many-body_2013}}. 
We also make comparison with dRPA+SOSEX, a variant of ph-RPA and second-order screened exchange common in condensed matter physics~\cite{eshuis_electron_2012,Note1}. %

The range separated scheme reported here returns an energy that is consistently within 0.001 a.u. per electron of the exact result, and is of comparable or better accuracy to dRPA+SOSEX energies and CCD. For more generality we also plot the same curve for the 54 electron system for which only diffusion Monte Carlo results are available~\cite{kwon_effects_1998,lopez_rios_inhomogeneous_2006}.
 Since these only have an accuracy of around 0.001 a.u. per electron in the high density regime~\cite{shepherd_full_2012}, the disagreement between the range separation energy and QMC is slightly larger.

For further comparison, we provide different CCD calculations where different channels have been excluded. 
The nomenclature in this paper uses prefixes to denote included channels in a calculation: r for rings, l for ladders, x for crossed-rings and m for mosaics. For example, rmCCD is a CCD calculation performed with the driving, rings and mosaic terms only.
The two methods shown, rmCCD (similar to ph-RPA) and lmCCD (similar to pp-RPA), are known to overcorrelate and undercorrelate respectively; their combination in range separated CCD performs significantly better.

\begin{figure}
    \subfloat[\label{fig2a}]{%
      \includegraphics[width=0.25\textwidth]{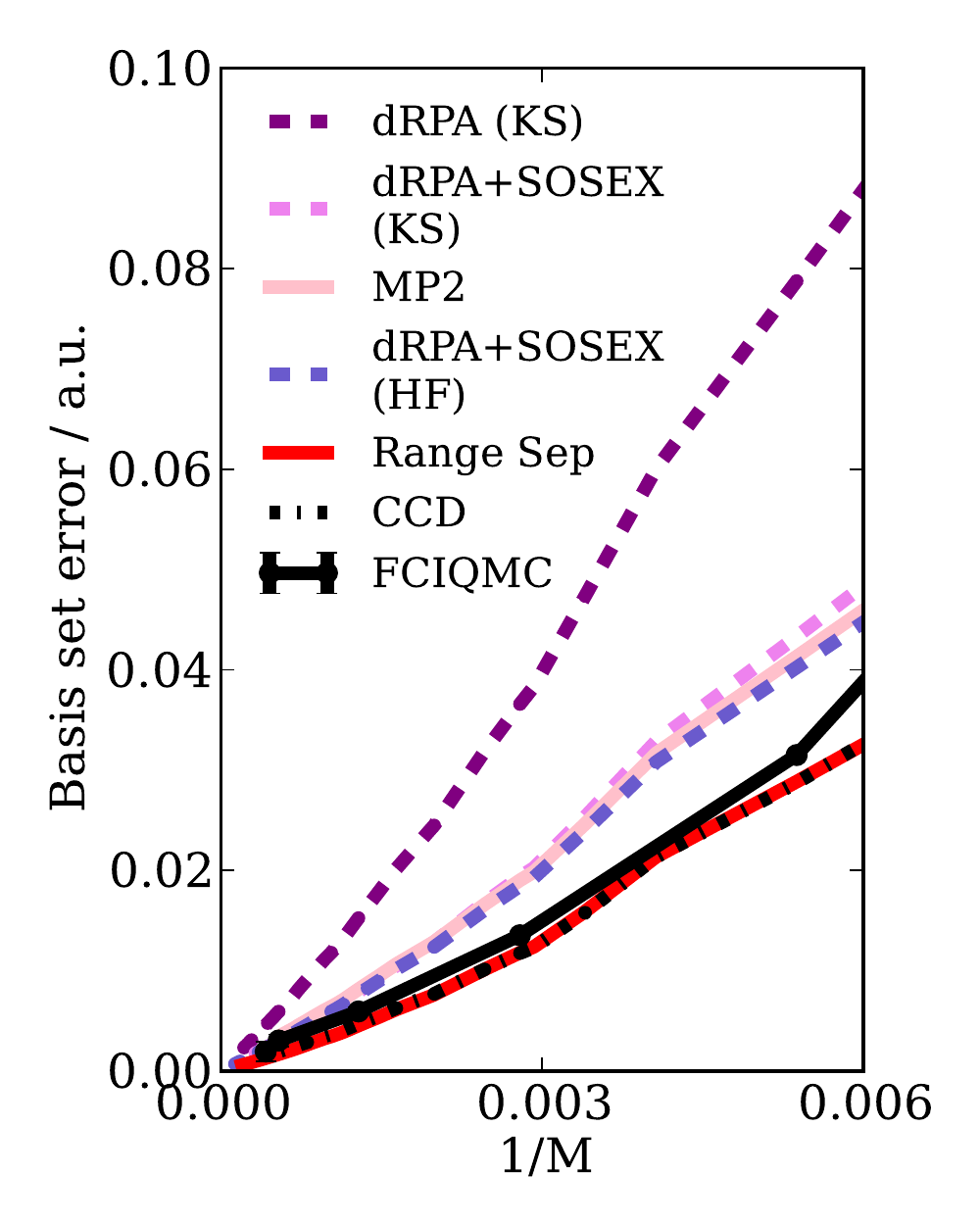}
          }
    \subfloat[\label{fig2b}]{%
      \includegraphics[width=0.25\textwidth]{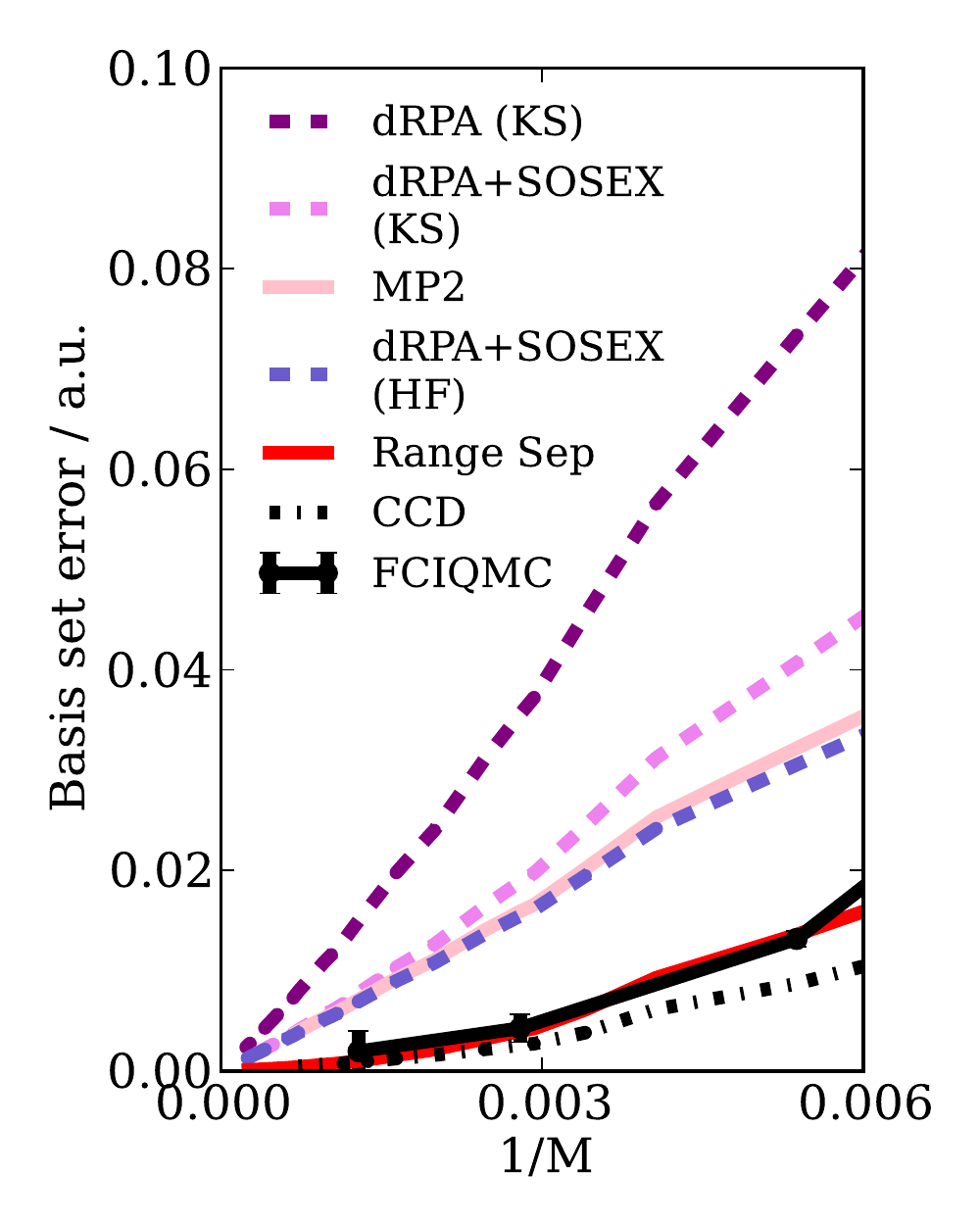}
          }
    \caption{Comparison of approach to the complete basis set limit between different methods at (a) $r_s=1.0$ and (b) $r_s=5.0$. Error bars on the QMC data are shown at 2$\sigma$, where they better represent 95\% confidence intervals, for emphasis. ($N$=14, $M\rightarrow\infty$ by extrapolation~\cite{shepherd_investigation_2012}).}
    \label{fig2}
\end{figure}

{\bf\emph{Basis set convergence.--} }
Although sometimes overlooked, an important aspect of basis set methods is the manner in which the correlation energy is captured as the basis set is expanded. 
Basis sets which are said to describe correlation consistently describe the physics incrementally and allow for the best extrapolation.
For a given Hamiltonian and basis set, this is related to the manner in which a method captures static (strong) versus dynamic (weak) correlation. 
Exact results from exact diagonalisation in a finite basis sets are the best benchmarks to compare behaviour between methods.

It has been remarked that dRPA+SOSEX based on a Kohn--Sham reference performs spuriously well for the electron gas~\cite{gruneis_making_2009}, in spite of the description of the correlation hole likely being similar to dRPA and hence too deep~\cite{singwi_electron_1968,kurth_density-functional_1999}. 
The overestimation of the correlation energy due to this comes from basis functions with high momenta, and can therefore be seen in the basis set extrapolation curves for large basis set sizes ($M$)~\cite{shepherd_convergence_2012}. This is shown in \reffig{fig2}, where a variety of methods are compared. Theories derived from dRPA and MP2 all have a behaviour that enters the linear $1/M$ regime too quickly and with too steep a gradient. The quality of the Kohn--Sham dRPA+SOSEX result does not therefore translate to good basis set energies, and therefore can be seen to result from a cancellation of capturing too little correlation energy around the Fermi surface and too much correlation energy at higher momenta. 

In contrast, our range-separated scheme retains the much improved behaviour of CCD.

\begin{figure*}
    \subfloat[\label{fig3}]{%
      \includegraphics[width=0.45\textwidth]{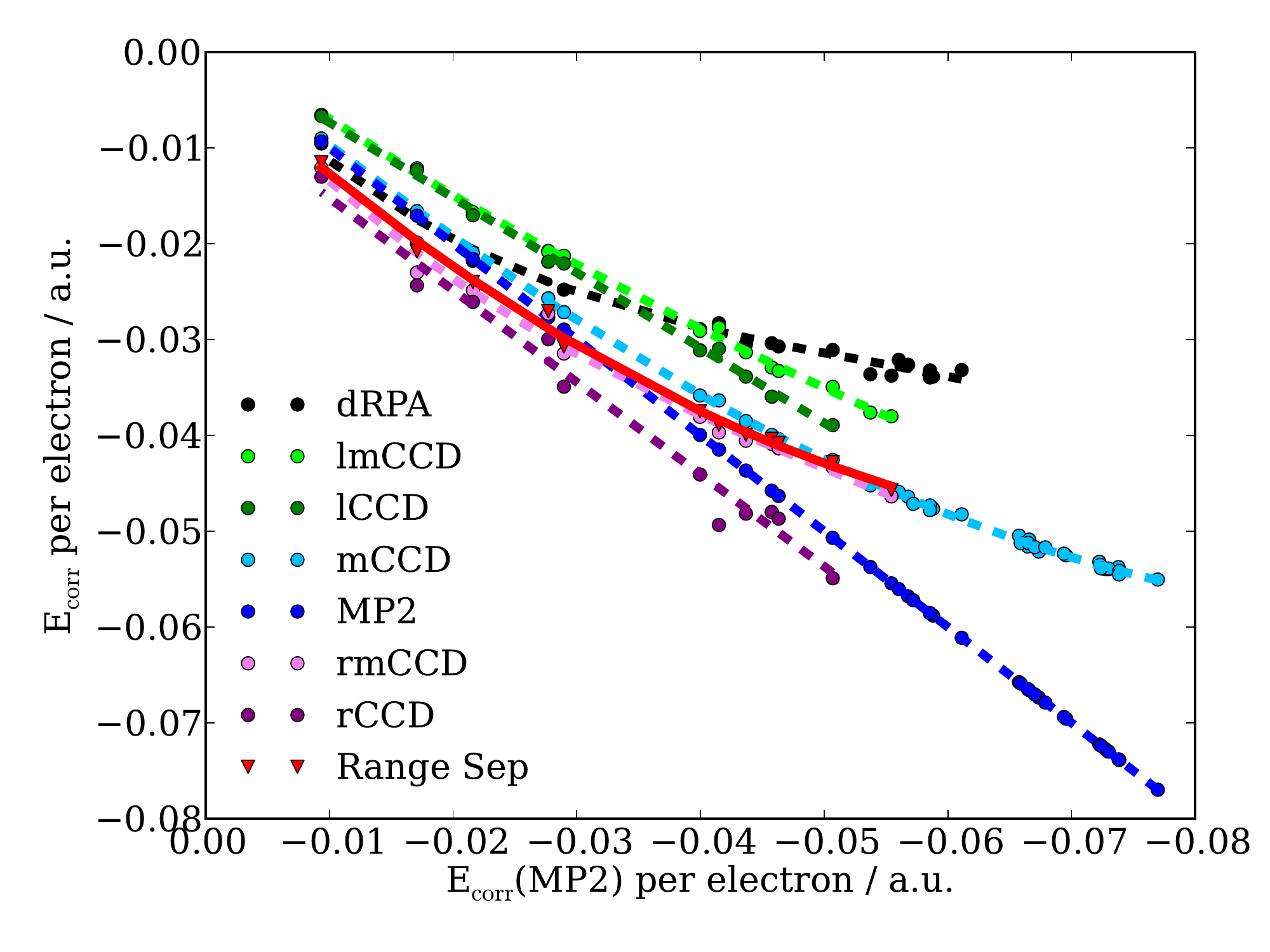}
          }
    \subfloat[\label{fig4}]{%
      \includegraphics[width=0.45\textwidth]{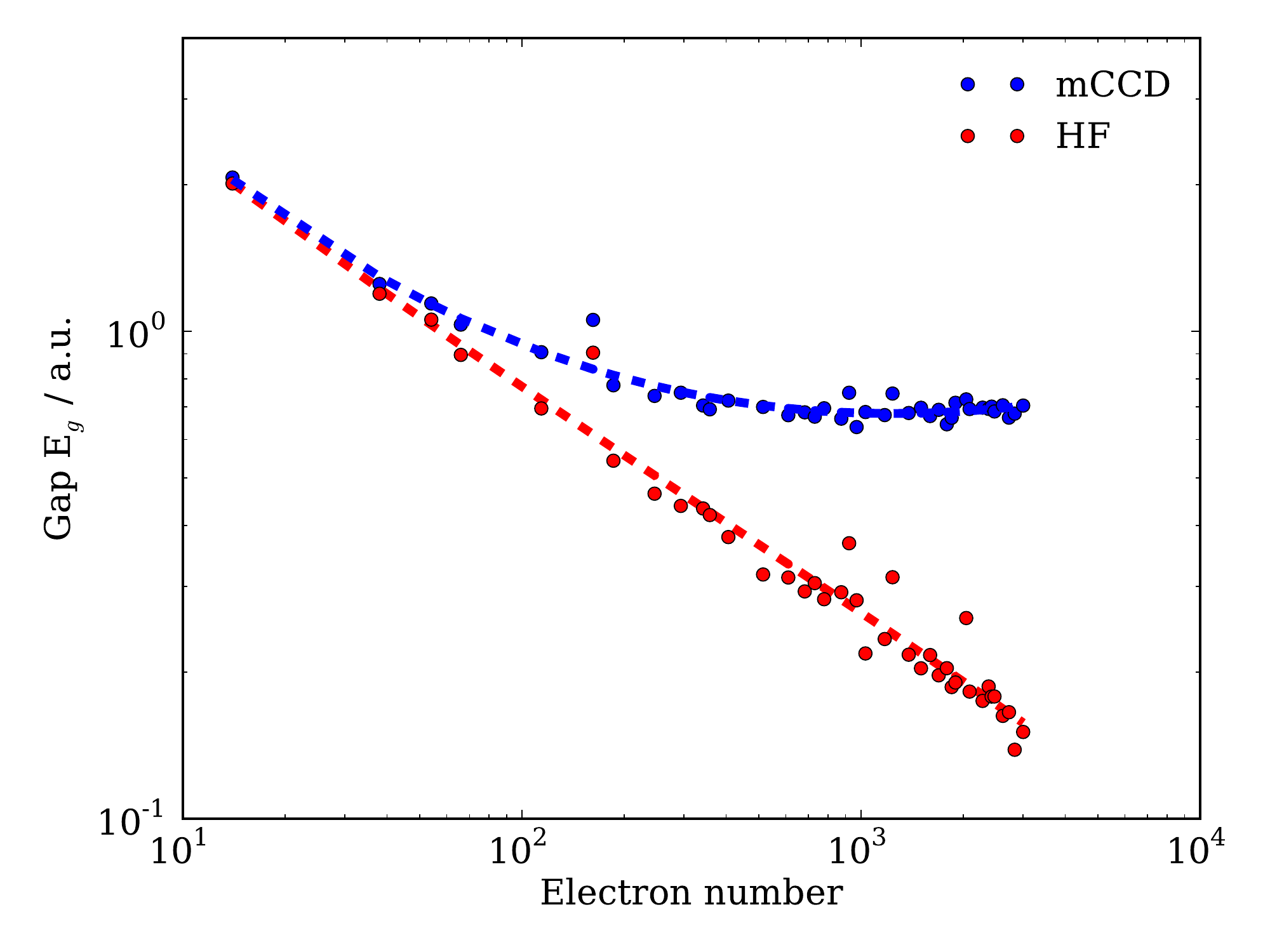}
          }
    \caption{Behaviour on approach to the TDL. Each point on this graph represents a single $N$ and $M$ and values vary between $N=14-3006$ with $M=38-8338$; the gas density was $r_s$=1.0~a.u.
In (a), we compare energies. Divergent methods appear as a straight line, apparent convergences as a curve. The lines only represent fits after the method is determined to converge or diverge, and are only intended as a guide to the eye.
In (b), we compare the orbital energy gap HF and the same gap in mCCD; the latter remains positive and finite in the TDL (see text).
    }
\end{figure*}

{\bf\emph{Thermodynamic limit.--} }
In the thermodynamic limit (TDL), finite order perturbation theories can yield divergent energies. One example of this is that the second-order MP2 energy diverges. 
There has been substantial success in examining these analytically, but with CC theories this can be very difficult due to the need to solve non-linear equations.
Approaching this limit numerically is also difficult due to slow divergences and the rising cost of simulating additional particles.

We follow a recent method due to Shepherd and Gr\"uneis~\cite{shepherd_many-body_2013}.
In outline, this explored a minimal representation of the area in $k$-space around the Fermi sphere which is the origin of the low-momentum excitations that cause the divergence.
This amounts to performing a numerical quadrature of a small area around the Fermi sphere (typically of radius $\sqrt{2}$ larger) by more and more finely spaced grids in such a way that the divergence can be seen by constant and predictable growth of the energy.
By a judicious choice of basis set sizes for each system size, computational cost can be controlled. 
Although only a small fraction of the total energy is represented in this region, it was shown to be enough to demonstrate the divergence in the energy.

Figure \ref{fig3} plots such an analysis for various methods described in this paper.
MP2 is well understood to have a divergent energy and can be used therefore as a divergence benchmark. 
We plot MP2 energies against energies obtained with different theories. A correlation with MP2 suggests that the method in question diverges at the same speed as MP2.
Since all higher order divergences are quicker, the appearance of a deviation from correlation with MP2 implies a convergent energy. This is exemplified by dRPA.
Our companion study discusses these limitations more thoroughly~\cite{shepherd_coupled_2013}.

As expected, the dRPA does not have a linear correlation with MP2, nor does CCD which would overlay the dRPA line on this scale. 
The ladder-only diagrams (lCCD) diverge, which is also consistent with a variety of previous comments and discussions in the literature (\emph{e.g.}~\cite{freeman_coupled-cluster_1983}). More surprisingly, the same appears true for rCCD which is subtly different from dRPA~\cite{scuseria_ground_2008}. In general, however, methods can be made to converge by the inclusion of mosaic diagrams and changing the reference for the calculation to the Brueckner Hamiltonian. This can be seen most dramatically by comparing pairs of lines with and without mosaics: rCCD with rmCCD; lCCD with lmCCD; and MP2 with mCCD. This test also shows that the range-separated scheme converges, as desired.

Finite electron gases have gapped orbital energy spectra due to the finite spacing of $k$-states.  The orbital energy gap calculated from Hartree-Fock theory closes in the TDL, which can be seen by examining the gap for a series of system sizes as shown in \reffig{fig4}. The closure of this orbital energy gap is one of the reasons that perturbation theories diverge for metals, since the smallest energy denominator is just twice the smallest orbital energy gap. In contrast, adding mosaic terms yields a modified single-particle spectrum which remains gapped even in the TDL, which suppresses the divergences, though we must point out that this non-zero orbital energy gap does not imply that the many-particle wave function is insulating. More discussion of this is included in the supplementary information~\cite{Note1}.  %

{\bf\emph{Conclusion.--} }
We have made an approximation of the CCD equations using ideas inspired from range separation in density functional theory and applied it to finite electron gas systems. We find a combination of terms which couple together pp-RPA in the short range and ph-RPA in the long range and in a manner that is superior to both methods.

This allows us to propose an approximation with the same computational cost scaling as CCD with several desirable properties: (a) improved accuracy over CCD, especially in the low-density regime; (b) improved accuracy over various forms of dRPA+SOSEX, especially in calculating energies from high-momentum basis functions; (c) retention of a finite energy per particle in the thermodynamic limit. This method performs well over a wide range of density regimes. %
Therefore, we hope that the method presented here will be transferable to real periodic and extended systems, where there is a growing interest in treating problems in solid state materials science with wavefunction and many-body methods.%

{\bf\emph{Acknowledgements.--} }
The authors thank Andreas Savin and Andreas Gr\"uneis for helpful discussions and the authors of Ref.~\onlinecite{shepherd__2013} for early access to their data. One of us (JJS) would like to thank the Royal Commission for the Exhibition of 1851 for a Research Fellowship. This work was supported by the National Science Foundation (CHE-1102601), the Welch Foundation (C-0036) and DOE-CMCSN (DE-SC0006650).

\cleardoublepage



\section*{Supplementary material}

# Supplementary Information for "Range Separated Brueckner Coupled Cluster Doubles Theory"

James J. Shepherd,[*] Thomas M. Henderson, and Gustavo E. Scuseria

*Department of Chemistry and Department of Physics and Astronomy, Rice University, Houston, TX 77005-1892*

(Dated: May 5, 2014)

### 1. Coupled cluster and dRPA

The CCD amplitude equations read, in the canonical Hartree–Fock basis,

$$\begin{aligned}(\epsilon_i + \epsilon_j - \epsilon_a - \epsilon_b)t_{ij}^{ab} &= \bar{v}_{ij}^{ab} \\ &+ \frac{1}{2}\bar{v}_{cd}^{ab}t_{ij}^{cd} + \frac{1}{2}\bar{v}_{ij}^{kl}t_{kl}^{ab} + \frac{1}{4}\bar{v}_{cd}^{kl}t_{ij}^{cd}t_{kl}^{ab} \\ &+ \bar{v}_{cj}^{kb}t_{ik}^{ac} + \bar{v}_{ci}^{ka}t_{jk}^{bc} + \bar{v}_{cd}^{kl}t_{lj}^{db}t_{ik}^{ac} \\ &- \bar{v}_{cj}^{ka}t_{ik}^{bc} - \bar{v}_{ci}^{kb}t_{jk}^{ac} - \bar{v}_{cd}^{kl}t_{lj}^{da}t_{ik}^{bc} \\ &+ \frac{1}{2}\bar{v}_{cd}^{kl}\left[t_{lj}^{ab}t_{ik}^{cd} - t_{li}^{ab}t_{jk}^{cd} + t_{ji}^{db}t_{kl}^{ac} - t_{ij}^{da}t_{kl}^{bc}\right]\end{aligned} \quad (1)$$

where in these equations: $\bar{v}_{ab}^{ij}$ is the two-body interaction in a spin-orbital basis using Dirac's notation, $\bar{v}_{ab}^{ij} = v_{ab}^{ij} - v_{ba}^{ij} = \langle ij||ab\rangle$; $\epsilon$ are the Hartree–Fock eigenvalues; and repeated indices are summed on the right-hand side. The indices $i, j, k,$ and $l$ label hole states; $a, b, c,$ and $d$ label particle states. This expression is grouped such that the top line represents the driver term, the second line ladder terms (l), the third line ring (r) terms, the fourth line crossed-ring (x) terms and the final line mosaics (m). The mosaic terms are so-called because they involve simultaneous ring and ladder contractions.

In writing the equations in this manner, we can consider solving each channel independently or in specific combinations; this has previously been of interest because the ring channel corresponds to particle-hole RPA [1] and the ladder channel to particle-particle RPA [2]. In the main paper, we use the nomenclature that rings-only CCD would be rCCD *etc.*

These non-linear equations are solved for amplitudes $t$ [3], from which the correlation energy is computed:

$$E_{\text{corr}} = \frac{1}{4}t_{ij}^{ab}\bar{v}_{ab}^{ij}. \quad (2)$$

Relating particle-hole RPA to rings-only CCD involves the removal of the antisymmetric integrals (*i.e.* exchange-like integrals) from the $\bar{v}_{ab}^{ij}$ terms in the amplitude equations, yielding [1]:

$$\begin{aligned}0 = &v_{ij}^{ab} + t_{ij}^{ab}(\epsilon_a + \epsilon_b - \epsilon_i - \epsilon_j) \\ &+ v_{cj}^{kb}t_{ik}^{ac} + v_{ci}^{ka}t_{jk}^{bc} + v_{cd}^{kl}t_{lj}^{db}t_{ik}^{ac}.\end{aligned} \quad (3)$$

A different energy expression is used,

$$E_{\text{corr}} = \frac{1}{2}t_{ij}^{ab}\bar{v}_{ab}^{ij}, \quad (4)$$

and this energy yields what is commonly called the dRPA+SOSEX energy. Removal of the SOSEX term, which involves removing the anti-symmetrisation from the expression above, yields the dRPA energy. The factor of a half comes from the plasmon formula for the dRPA energy. It is then possible to obtain both of these energies using a Hartree–Fock or a Kohn–Sham determinant reference. For the electron gas the latter is equivalent to using a non-interacting reference because the constant contribution from the exchange-correlation potential cancels when energy differences are taken.

### 2. Range separation

It is possible to partition the Coulomb potential using an exponential function:

$$\frac{1}{r_{12}} = \frac{e^{-\gamma r_{12}}}{r_{12}} + \frac{(1 - e^{-\gamma r_{12}})}{r_{12}}. \quad (5)$$

This corresponds to a Yukawa potential in the short range and as such a reasonable choice of $\gamma$ should be the Thomas-Fermi screening parameter. The Thomas–Fermi screening parameter is a reasonable choice for the electron gas, being also the static limit of the screening within the RPA and is given by,

$$\gamma^2 = \frac{4}{\pi}k_F, \quad (6)$$

where $k_F$ is the Fermi wave vector. Overall therefore this is a density-dependent screening.

### 3. Range separated coupled cluster

Starting with mCCD, we can incorporate the mosaic terms into the one-particle eigenvalues:

$$(\eta_i + \eta_j - \eta_a - \eta_b)t_{ij}^{ab} = \bar{v}_{ij}^{ab}. \quad (7)$$

Here, the new eigenvalues $\eta$ are given by:

$$\eta_i = \epsilon_i + \frac{1}{2}\bar{v}_{cd}^{il}t_{il}^{cd} \quad (8)$$

$$\eta_a = \epsilon_a - \frac{1}{2}\bar{v}_{ad}^{kl}t_{kl}^{ad}, \quad (9)$$

---

[*] jjs6@rice.edu

and form what can be called the Brueckner one-body Hamiltonian [2, 4, 5]. We have made use of this being diagonal in a canonical plane wave basis set due to momentum conservation.

We can now add in the rings using a long-range potential and add the ladder terms with a short-range potential:

$$(\eta_i + \eta_j - \eta_a - \eta_b)t_{ij}^{ab} = \bar{v}_{ij}^{ab}$$
$$+ \frac{1}{2}(\bar{v}_{\mathrm{SR}})_{cd}^{ab}t_{ij}^{cd} + \frac{1}{2}(\bar{v}_{\mathrm{SR}})_{ij}^{kl}t_{kl}^{ab} + \frac{1}{4}(\bar{v}_{\mathrm{SR}})_{cd}^{kl}t_{ij}^{cd}t_{kl}^{ab} \quad (10)$$
$$+ (\bar{v}_{\mathrm{LR}})_{cj}^{kb}t_{ik}^{ac} + (\bar{v}_{\mathrm{LR}})_{ci}^{ka}t_{jk}^{bc} + (\bar{v}_{\mathrm{LR}})_{cd}^{kl}t_{lj}^{db}t_{ik}^{ac}.$$

The crossed-rings are not present in these equations because they do not represent an harmonic oscillator problem that can be added into pp-RPA or ph-RPA [2].

### 4. Electron gas

We take as our model system the uniform electron gas:

$$\hat{H} = T + V_{\mathrm{ee}} + V_{\mathrm{eb}} + V_{\mathrm{bb}} \quad (11)$$

where these terms are the kinetic energy operator for the electrons, the electron-electron interaction, the electron-background interaction and the background-background interaction.

We work with a finite electron gas, where there are $N$ particles in a box of finite length $L$. The infinite particle, or thermodynamic, limit can be found by taking $N \to \infty$ whilst keeping the density constant.

Important ingredients required for our calculations include an assumed dimensionality and symmetry. Here, we use a 3D cubic gas for which the density is measured by the Wigner-Seitz radius:

$$\frac{4}{3}\pi r_s^3 = \frac{L^3}{N}. \quad (12)$$

We use a plane wave basis set,

$$\phi_j = \sqrt{\frac{1}{L^3}} \exp\left(i\mathbf{k}_j \cdot \mathbf{r}\right), \quad (13)$$

where $L$ is the length of the box. The momentum $\mathbf{k}$ is quantised:

$$\mathbf{k} = \frac{2\pi}{L}\mathbf{n} \quad (14)$$

where $\mathbf{n}$ is an integer vector. As the thermodynamic limit (TDL) is approached $L \to \infty$ and the spacing tends to zero.

The four-index integrals in a plane wave basis are,

$$v_{ab}^{ij} = \langle ij|V|ab\rangle \quad (15)$$
$$= v(\mathbf{k}_a - \mathbf{k}_i)\delta_{\mathbf{k}_a - \mathbf{k}_i, \mathbf{k}_j - \mathbf{k}_b}, \quad (16)$$
$$v(\mathbf{q}) = \begin{cases} \frac{1}{L^3}\frac{4\pi}{\mathbf{q}^2}, & \mathbf{q} \neq \mathbf{0} \\ v_M, & \mathbf{q} = \mathbf{0}. \end{cases} \quad (17)$$

and the Hartree–Fock eigenvalues are:

$$\epsilon_p = \frac{1}{2}\mathbf{k}_p^2 - \sum_{j \in \mathrm{occ}} v(\mathbf{k}_p - \mathbf{k}_j). \quad (18)$$

Here, momentum is conserved by the Kronecker delta:

$$\mathbf{k}_a - \mathbf{k}_i = \mathbf{k}_j - \mathbf{k}_b. \quad (19)$$

The zero momentum integral ($\mathbf{k}_a = \mathbf{k}_i$) is given by $v_M$, the Madelung constant, which is a feature of the finite electron gas which vanishes in the TDL found as $N \to \infty$ and $L \to \infty$. The Madelung constant is defined uniquely for a cell geometry and box length, and here we calculate it to be $v_M L \approx 2.8372$ where $L$ is the length of the box [6, 7].

For short-range integrals, the matrix elements are also easy to compute:

$$(v_{\mathrm{SR}})_{ab}^{ij} = \frac{1}{L^3}\frac{4\pi}{(\mathbf{q}^2 + \gamma^2)}. \quad (20)$$

The long-range integrals are then simply:

$$(v_{\mathrm{LR}})_{ab}^{ij} = v_{ij}^{ab} - (v_{\mathrm{SR}})_{ab}^{ij}. \quad (21)$$

A spherically symmetric kinetic energy cutoff is used to define the basis set with $M$ spin orbitals and for small systems FCI-quality benchmark data are available for comparison [8]. The energy approaches the complete basis set limit as $1/M$ to lead order [9].

### 5. Benchmarks

Since the Ceperley-Alder calculations [10], highly accurate results have been available for various electron gas systems both at finite $N$ and at the TDL. This has allowed modern advanced in diffusion Monte Carlo to focus on benchmarking finite electron gases [11, 12], which is sufficiently representative of real (periodic) systems.

In a series of recent papers, the electron gas has been explored from a quantum chemical perspective [8, 9, 13–15], adding to decades of work from condensed matter physicists. This has provided a number of technical advances making this work possible and in particular has allowed for the availability of FCI-quality finite basis benchmarks at a wide range of basis set sizes. This enables finite basis methods to be compared without the need of going to very large basis set sizes.

The set of benchmarks we have used for this study comes from full configuration interaction quantum Monte Carlo [14, 16] for the high-to-metallic density regime ($N = 14$), and diffusion Monte Carlo for the remainder [7, 11, 17] and are effectively exact. These results will be reported in more detail in upcoming work [18].

### 6. Divergences and energy gaps

The second-order Møller-Plesset energy is given by:

$$E_{\text{corr}} = \frac{1}{4} t_{ij}^{ab} \bar{v}_{ab}^{ij}, \qquad (22)$$

where the amplitudes can be written in terms of Hartree–Fock orbital energies:

$$t_{ij}^{ab} = \frac{\bar{v}_{ij}^{ab}}{\epsilon_i^{\text{HF}} + \epsilon_j^{\text{HF}} - \epsilon_a^{\text{HF}} - \epsilon_b^{\text{HF}}}. \qquad (23)$$

In the TDL, some of these terms have analytical expressions. The overall divergence is caused by a term that behaves as:

$$\int_{q=0} \frac{1}{q} dq, \qquad (24)$$

and although formal textbook derivations neglect exchange [19], modern numerical data show consistent results [15, 20].

An important contribution to this divergence is the vanishing of the minimum orbital energy gap:

$$\min(\epsilon_i^{\text{HF}} + \epsilon_j^{\text{HF}} - \epsilon_a^{\text{HF}} - \epsilon_b^{\text{HF}}) \to 0, \qquad (25)$$

around the Fermi energy. However, this is never observed directly in finite systems due to quantisation of the spectrum where the eigenvalue equations read:

$$\epsilon_i = \frac{1}{2} \mathbf{k}_i^2 - \sum_{j \in \text{occ}} v_{ij}^{ji}, \qquad (26)$$

and values of $\mathbf{k}$ are constrained by Eq. (14). In the TDL, the Hartree–Fock band structure can be found analytically [21]:

$$\epsilon(\mathbf{k}) = \frac{1}{2} \mathbf{k}^2 + \frac{k_F}{\pi} f(x) \qquad (27)$$

with

$$f(x) = -\left(1 + \frac{1-x^2}{2x} \ln\left|\frac{1+x}{1-x}\right|\right) \qquad (28)$$

and $x = |\mathbf{k}|/k_F$.

The physical interpretation of these minimum orbital energy gaps is that they are estimates for the band gap, which is assured by Koopman's theorem. The finite picture always has a finite gap and is an insulator, whereas in the thermodynamic limit the Hartree–Fock band structure predicts a zero gap semiconductor due to a zero in the density of states.

The Brueckner one-body Hamiltonian has a modified spectrum given in Eqs. 8 and 9. Since $v$ and $t$ have opposite signs for the electron gas, the occupied manifold will be pushed down in energy and the virtual manifold pushed up. The result is an increased minimum orbital energy gap. We present numerical evidence in the body of the paper that this gap persists to the TDL.

Is the Brueckner one-body Hamiltonian an insulating reference? In some respect yes: the gap remains open in the TDL. However, unlike Hartree–Fock theory there is no direct interpretation relating orbital energy differences to a physical band gap; we have no equivalent of Koopman's theorem for this effective Hamiltonian. It is perhaps better to view these as 'dressed' eigenvalues where some effects of correlation, and therefore screening, have been incorporated.

It is worthwhile noting that the characterisation of these states as metallic or insulating refers to the single determinant references and not to the correlated wave functions formed from them. In particular, examining the effective Brueckner one-body Hamiltonian does not give us information about the many-body wavefunction from mosaic CCD calculations.

[1] G. E. Scuseria, T. M. Henderson, and D. C. Sorensen, J. Chem. Phys. **129**, 231101 (2008).
[2] G. E. Scuseria, T. M. Henderson, and I. W. Bulik, J. Chem. Phys. **139**, 104113 (2013).
[3] G. E. Scuseria, T. J. Lee, and H. F. Schaefer III, Chem. Phys. Lett. **130**, 236 (1986).
[4] G. E. Scuseria, Int. J. Quant. Chem. **55**, 165–171 (1995).
[5] R. J. Bartlett, Chem. Phys. Lett. **484**, 1 (2009).
[6] L. M. Fraser, W. M. C. Foulkes, G. Rajagopal, R. J. Needs, S. D. Kenny, and A. J. Williamson, Phys. Rev. B **53**, 1814 (1996).
[7] R. J. Needs, M. D. Towler, N. D. Drummond, and P. López Ríos, J. Phys.: Condens. Matter **22**, 023201 (2010).
[8] J. J. Shepherd, G. Booth, A. Grüneis, and A. Alavi, Phys. Rev. B **85**, 081103 (2012).
[9] J. J. Shepherd, A. Grüneis, G. H. Booth, G. Kresse, and A. Alavi, Phys. Rev. B **86**, 035111 (2012).
[10] D. M. Ceperley and B. J. Alder, Phys. Rev. Lett. **45**, 566 (1980).
[11] P. López Ríos, A. Ma, N. D. Drummond, M. D. Towler, and R. J. Needs, Phys. Rev. E **74**, 066701 (2006).
[12] M. Holzmann, D. M. Ceperley, C. Pierleoni, and K. Esler, Phys. Rev. E **68**, 046707 (2003).
[13] A. Grüneis, J. J. Shepherd, A. Alavi, D. P. Tew, and G. H. Booth, J. Chem. Phys. **139**, 084112 (2013).
[14] J. J. Shepherd, G. H. Booth, and A. Alavi, J. Chem. Phys. **136**, 244101 (2012).

[15] J. J. Shepherd and A. Grüneis, Phys. Rev. Lett. **110**, 226401 (2013).
[16] G. H. Booth, A. Grüneis, G. Kresse, and A. Alavi, Nature **493**, 365 (2013).
[17] P. López Ríos, Pers. Comm. (2013).
[18] J. J. Shepherd, P. L. Ríos, N. D. Drummond, R. J. Needs, and A. Alavi, In preparation (2013).
[19] R. D. Mattuck, *A guide to Feynman diagrams in the many-body problem*, 2nd ed., Dover books on physics and chemistry (Dover Publications, New York, 1992).
[20] A. Grüneis, M. Marsman, and G. Kresse, J. Chem. Phys. **133**, 074107 (2010).
[21] R. M. Martin, *Electronic structure: basic theory and practical methods* (Cambridge University Press, Cambridge UK, 2004).




\begin{thebibliography}{55}%
\makeatletter
\providecommand \@ifxundefined [1]{%
 \@ifx{#1\undefined}
}%
\providecommand \@ifnum [1]{%
 \ifnum #1\expandafter \@firstoftwo
 \else \expandafter \@secondoftwo
 \fi
}%
\providecommand \@ifx [1]{%
 \ifx #1\expandafter \@firstoftwo
 \else \expandafter \@secondoftwo
 \fi
}%
\providecommand \natexlab [1]{#1}%
\providecommand \enquote  [1]{``#1''}%
\providecommand \bibnamefont  [1]{#1}%
\providecommand \bibfnamefont [1]{#1}%
\providecommand \citenamefont [1]{#1}%
\providecommand \href@noop [0]{\@secondoftwo}%
\providecommand \href [0]{\begingroup \@sanitize@url \@href}%
\providecommand \@href[1]{\@@startlink{#1}\@@href}%
\providecommand \@@href[1]{\endgroup#1\@@endlink}%
\providecommand \@sanitize@url [0]{\catcode `\\12\catcode `\$12\catcode
  `\&12\catcode `\#12\catcode `\^12\catcode `\_12\catcode `\%12\relax}%
\providecommand \@@startlink[1]{}%
\providecommand \@@endlink[0]{}%
\providecommand \url  [0]{\begingroup\@sanitize@url \@url }%
\providecommand \@url [1]{\endgroup\@href {#1}{\urlprefix }}%
\providecommand \urlprefix  [0]{URL }%
\providecommand \Eprint [0]{\href }%
\providecommand \doibase [0]{http://dx.doi.org/}%
\providecommand \selectlanguage [0]{\@gobble}%
\providecommand \bibinfo  [0]{\@secondoftwo}%
\providecommand \bibfield  [0]{\@secondoftwo}%
\providecommand \translation [1]{[#1]}%
\providecommand \BibitemOpen [0]{}%
\providecommand \bibitemStop [0]{}%
\providecommand \bibitemNoStop [0]{.\EOS\space}%
\providecommand \EOS [0]{\spacefactor3000\relax}%
\providecommand \BibitemShut  [1]{\csname bibitem#1\endcsname}%
\let\auto@bib@innerbib\@empty
%
\bibitem [{\citenamefont {Bohm}\ and\ \citenamefont
  {Pines}(1951)}]{bohm_collective_1951}%
  \BibitemOpen
  \bibfield  {author} {\bibinfo {author} {\bibfnamefont {D.}~\bibnamefont
  {Bohm}}\ and\ \bibinfo {author} {\bibfnamefont {D.}~\bibnamefont {Pines}},\
  }\href {\doibase 10.1103/PhysRev.82.625} {\bibfield  {journal} {\bibinfo
  {journal} {Phys. Rev.}\ }\textbf {\bibinfo {volume} {82}},\ \bibinfo {pages}
  {625} (\bibinfo {year} {1951})}\BibitemShut {NoStop}%
\bibitem [{\citenamefont {Pines}\ and\ \citenamefont
  {Bohm}(1952)}]{pines_collective_1952}%
  \BibitemOpen
  \bibfield  {author} {\bibinfo {author} {\bibfnamefont {D.}~\bibnamefont
  {Pines}}\ and\ \bibinfo {author} {\bibfnamefont {D.}~\bibnamefont {Bohm}},\
  }\href {\doibase 10.1103/PhysRev.85.338} {\bibfield  {journal} {\bibinfo
  {journal} {Phys. Rev.}\ }\textbf {\bibinfo {volume} {85}},\ \bibinfo {pages}
  {338} (\bibinfo {year} {1952})}\BibitemShut {NoStop}%
\bibitem [{\citenamefont {Bohm}\ and\ \citenamefont
  {Pines}(1953)}]{bohm_collective_1953}%
  \BibitemOpen
  \bibfield  {author} {\bibinfo {author} {\bibfnamefont {D.}~\bibnamefont
  {Bohm}}\ and\ \bibinfo {author} {\bibfnamefont {D.}~\bibnamefont {Pines}},\
  }\href {\doibase 10.1103/PhysRev.92.609} {\bibfield  {journal} {\bibinfo
  {journal} {Phys. Rev.}\ }\textbf {\bibinfo {volume} {92}},\ \bibinfo {pages}
  {609} (\bibinfo {year} {1953})}\BibitemShut {NoStop}%
\bibitem [{\citenamefont {Harl}\ and\ \citenamefont
  {Kresse}(2009)}]{harl_accurate_2009}%
  \BibitemOpen
  \bibfield  {author} {\bibinfo {author} {\bibfnamefont {J.}~\bibnamefont
  {Harl}}\ and\ \bibinfo {author} {\bibfnamefont {G.}~\bibnamefont {Kresse}},\
  }\href {\doibase 10.1103/PhysRevLett.103.056401} {\bibfield  {journal}
  {\bibinfo  {journal} {Phys. Rev. Lett.}\ }\textbf {\bibinfo {volume} {103}},\
  \bibinfo {pages} {056401} (\bibinfo {year} {2009})}\BibitemShut {NoStop}%
\bibitem [{\citenamefont {Leininger}\ \emph {et~al.}(1997)\citenamefont
  {Leininger}, \citenamefont {Stoll}, \citenamefont {Werner},\ and\
  \citenamefont {Savin}}]{leininger_combining_1997}%
  \BibitemOpen
  \bibfield  {author} {\bibinfo {author} {\bibfnamefont {T.}~\bibnamefont
  {Leininger}}, \bibinfo {author} {\bibfnamefont {H.}~\bibnamefont {Stoll}},
  \bibinfo {author} {\bibfnamefont {H.-J.}\ \bibnamefont {Werner}}, \ and\
  \bibinfo {author} {\bibfnamefont {A.}~\bibnamefont {Savin}},\ }\href
  {\doibase 10.1016/S0009-2614(97)00758-6} {\bibfield  {journal} {\bibinfo
  {journal} {Chem. Phys. Lett.}\ }\textbf {\bibinfo {volume} {275}},\ \bibinfo
  {pages} {151} (\bibinfo {year} {1997})}\BibitemShut {NoStop}%
\bibitem [{\citenamefont {Goll}\ \emph {et~al.}(2005)\citenamefont {Goll},
  \citenamefont {Werner},\ and\ \citenamefont {Stoll}}]{goll_short-range_2005}%
  \BibitemOpen
  \bibfield  {author} {\bibinfo {author} {\bibfnamefont {E.}~\bibnamefont
  {Goll}}, \bibinfo {author} {\bibfnamefont {H.-J.}\ \bibnamefont {Werner}}, \
  and\ \bibinfo {author} {\bibfnamefont {H.}~\bibnamefont {Stoll}},\ }\href
  {\doibase 10.1039/B509242F} {\bibfield  {journal} {\bibinfo  {journal} {Phys.
  Chem. Chem. Phys.}\ }\textbf {\bibinfo {volume} {7}},\ \bibinfo {pages}
  {3917} (\bibinfo {year} {2005})}\BibitemShut {NoStop}%
\bibitem [{\citenamefont {Goll}\ \emph {et~al.}(2006)\citenamefont {Goll},
  \citenamefont {Werner}, \citenamefont {Stoll}, \citenamefont {Leininger},
  \citenamefont {Gori-Giorgi},\ and\ \citenamefont
  {Savin}}]{goll_short-range_2006}%
  \BibitemOpen
  \bibfield  {author} {\bibinfo {author} {\bibfnamefont {E.}~\bibnamefont
  {Goll}}, \bibinfo {author} {\bibfnamefont {H.-J.}\ \bibnamefont {Werner}},
  \bibinfo {author} {\bibfnamefont {H.}~\bibnamefont {Stoll}}, \bibinfo
  {author} {\bibfnamefont {T.}~\bibnamefont {Leininger}}, \bibinfo {author}
  {\bibfnamefont {P.}~\bibnamefont {Gori-Giorgi}}, \ and\ \bibinfo {author}
  {\bibfnamefont {A.}~\bibnamefont {Savin}},\ }\href {\doibase
  10.1016/j.chemphys.2006.05.020} {\bibfield  {journal} {\bibinfo  {journal}
  {Chemical Physics}\ }\textbf {\bibinfo {volume} {329}},\ \bibinfo {pages}
  {276} (\bibinfo {year} {2006})}\BibitemShut {NoStop}%
\bibitem [{\citenamefont {Leb{\`e}gue}\ \emph {et~al.}(2010)\citenamefont
  {Leb{\`e}gue}, \citenamefont {Harl}, \citenamefont {Gould}, \citenamefont
  {{\'A}ngy{\'a}n}, \citenamefont {Kresse},\ and\ \citenamefont
  {Dobson}}]{lebegue_cohesive_2010}%
  \BibitemOpen
  \bibfield  {author} {\bibinfo {author} {\bibfnamefont {S.}~\bibnamefont
  {Leb{\`e}gue}}, \bibinfo {author} {\bibfnamefont {J.}~\bibnamefont {Harl}},
  \bibinfo {author} {\bibfnamefont {T.}~\bibnamefont {Gould}}, \bibinfo
  {author} {\bibfnamefont {J.~G.}\ \bibnamefont {{\'A}ngy{\'a}n}}, \bibinfo
  {author} {\bibfnamefont {G.}~\bibnamefont {Kresse}}, \ and\ \bibinfo {author}
  {\bibfnamefont {J.~F.}\ \bibnamefont {Dobson}},\ }\href {\doibase
  10.1103/PhysRevLett.105.196401} {\bibfield  {journal} {\bibinfo  {journal}
  {Phys. Rev. Lett.}\ }\textbf {\bibinfo {volume} {105}},\ \bibinfo {pages}
  {196401} (\bibinfo {year} {2010})}\BibitemShut {NoStop}%
\bibitem [{\citenamefont {Dobson}\ \emph {et~al.}(2006)\citenamefont {Dobson},
  \citenamefont {White},\ and\ \citenamefont
  {Rubio}}]{dobson_asymptotics_2006}%
  \BibitemOpen
  \bibfield  {author} {\bibinfo {author} {\bibfnamefont {J.~F.}\ \bibnamefont
  {Dobson}}, \bibinfo {author} {\bibfnamefont {A.}~\bibnamefont {White}}, \
  and\ \bibinfo {author} {\bibfnamefont {A.}~\bibnamefont {Rubio}},\ }\href
  {\doibase 10.1103/PhysRevLett.96.073201} {\bibfield  {journal} {\bibinfo
  {journal} {Phys. Rev. Lett.}\ }\textbf {\bibinfo {volume} {96}},\ \bibinfo
  {pages} {073201} (\bibinfo {year} {2006})}\BibitemShut {NoStop}%
\bibitem [{\citenamefont {Toulouse}\ \emph {et~al.}(2009)\citenamefont
  {Toulouse}, \citenamefont {Gerber}, \citenamefont {Jansen}, \citenamefont
  {Savin},\ and\ \citenamefont
  {{\'A}ngy{\'a}n}}]{toulouse_adiabatic-connection_2009}%
  \BibitemOpen
  \bibfield  {author} {\bibinfo {author} {\bibfnamefont {J.}~\bibnamefont
  {Toulouse}}, \bibinfo {author} {\bibfnamefont {I.~C.}\ \bibnamefont
  {Gerber}}, \bibinfo {author} {\bibfnamefont {G.}~\bibnamefont {Jansen}},
  \bibinfo {author} {\bibfnamefont {A.}~\bibnamefont {Savin}}, \ and\ \bibinfo
  {author} {\bibfnamefont {J.~G.}\ \bibnamefont {{\'A}ngy{\'a}n}},\ }\href
  {\doibase 10.1103/PhysRevLett.102.096404} {\bibfield  {journal} {\bibinfo
  {journal} {Phys. Rev. Lett.}\ }\textbf {\bibinfo {volume} {102}},\ \bibinfo
  {pages} {096404} (\bibinfo {year} {2009})}\BibitemShut {NoStop}%
\bibitem [{\citenamefont {Janesko}\ \emph {et~al.}(2009)\citenamefont
  {Janesko}, \citenamefont {Henderson},\ and\ \citenamefont
  {Scuseria}}]{janesko_long-range-corrected_2009}%
  \BibitemOpen
  \bibfield  {author} {\bibinfo {author} {\bibfnamefont {B.~G.}\ \bibnamefont
  {Janesko}}, \bibinfo {author} {\bibfnamefont {T.~M.}\ \bibnamefont
  {Henderson}}, \ and\ \bibinfo {author} {\bibfnamefont {G.~E.}\ \bibnamefont
  {Scuseria}},\ }\href {\doibase doi:10.1063/1.3090814} {\bibfield  {journal}
  {\bibinfo  {journal} {J. Chem. Phys.}\ }\textbf {\bibinfo {volume} {130}},\
  \bibinfo {pages} {081105} (\bibinfo {year} {2009})}\BibitemShut {NoStop}%
\bibitem [{\citenamefont {Bruneval}(2012)}]{bruneval_range-separated_2012}%
  \BibitemOpen
  \bibfield  {author} {\bibinfo {author} {\bibfnamefont {F.}~\bibnamefont
  {Bruneval}},\ }\href {\doibase 10.1103/PhysRevLett.108.256403} {\bibfield
  {journal} {\bibinfo  {journal} {Phys. Rev. Lett.}\ }\textbf {\bibinfo
  {volume} {108}},\ \bibinfo {pages} {256403} (\bibinfo {year}
  {2012})}\BibitemShut {NoStop}%
\bibitem [{\citenamefont {Singwi}\ \emph {et~al.}(1968)\citenamefont {Singwi},
  \citenamefont {Tosi}, \citenamefont {Land},\ and\ \citenamefont
  {Sj{\"o}lander}}]{singwi_electron_1968}%
  \BibitemOpen
  \bibfield  {author} {\bibinfo {author} {\bibfnamefont {K.~S.}\ \bibnamefont
  {Singwi}}, \bibinfo {author} {\bibfnamefont {M.~P.}\ \bibnamefont {Tosi}},
  \bibinfo {author} {\bibfnamefont {R.~H.}\ \bibnamefont {Land}}, \ and\
  \bibinfo {author} {\bibfnamefont {A.}~\bibnamefont {Sj{\"o}lander}},\ }\href
  {\doibase 10.1103/PhysRev.176.589} {\bibfield  {journal} {\bibinfo  {journal}
  {Phys. Rev.}\ }\textbf {\bibinfo {volume} {176}},\ \bibinfo {pages} {589}
  (\bibinfo {year} {1968})}\BibitemShut {NoStop}%
\bibitem [{\citenamefont {Kurth}\ and\ \citenamefont
  {Perdew}(1999)}]{kurth_density-functional_1999}%
  \BibitemOpen
  \bibfield  {author} {\bibinfo {author} {\bibfnamefont {S.}~\bibnamefont
  {Kurth}}\ and\ \bibinfo {author} {\bibfnamefont {J.~P.}\ \bibnamefont
  {Perdew}},\ }\href {\doibase 10.1103/PhysRevB.59.10461} {\bibfield  {journal}
  {\bibinfo  {journal} {Phys. Rev. B}\ }\textbf {\bibinfo {volume} {59}},\
  \bibinfo {pages} {10461} (\bibinfo {year} {1999})}\BibitemShut {NoStop}%
\bibitem [{\citenamefont {Freeman}(1977)}]{freeman_coupled-cluster_1977}%
  \BibitemOpen
  \bibfield  {author} {\bibinfo {author} {\bibfnamefont {D.~L.}\ \bibnamefont
  {Freeman}},\ }\href {\doibase 10.1103/PhysRevB.15.5512} {\bibfield  {journal}
  {\bibinfo  {journal} {Phys. Rev. B}\ }\textbf {\bibinfo {volume} {15}},\
  \bibinfo {pages} {5512} (\bibinfo {year} {1977})}\BibitemShut {NoStop}%
\bibitem [{\citenamefont {Gell-Mann}\ and\ \citenamefont
  {Brueckner}(1957)}]{gell-mann_correlation_1957}%
  \BibitemOpen
  \bibfield  {author} {\bibinfo {author} {\bibfnamefont {M.}~\bibnamefont
  {Gell-Mann}}\ and\ \bibinfo {author} {\bibfnamefont {K.~A.}\ \bibnamefont
  {Brueckner}},\ }\href {\doibase 10.1103/PhysRev.106.364} {\bibfield
  {journal} {\bibinfo  {journal} {Phys. Rev.}\ }\textbf {\bibinfo {volume}
  {106}},\ \bibinfo {pages} {364} (\bibinfo {year} {1957})}\BibitemShut
  {NoStop}%
\bibitem [{\citenamefont {Eshuis}\ \emph {et~al.}(2012)\citenamefont {Eshuis},
  \citenamefont {Bates},\ and\ \citenamefont {Furche}}]{eshuis_electron_2012}%
  \BibitemOpen
  \bibfield  {author} {\bibinfo {author} {\bibfnamefont {H.}~\bibnamefont
  {Eshuis}}, \bibinfo {author} {\bibfnamefont {J.~E.}\ \bibnamefont {Bates}}, \
  and\ \bibinfo {author} {\bibfnamefont {F.}~\bibnamefont {Furche}},\ }\href
  {\doibase 10.1007/s00214-011-1084-8} {\bibfield  {journal} {\bibinfo
  {journal} {Theor. Chem. Acc.}\ }\textbf {\bibinfo {volume} {131}},\ \bibinfo
  {pages} {1} (\bibinfo {year} {2012})}\BibitemShut {NoStop}%
\bibitem [{\citenamefont {Paier}\ \emph {et~al.}(2012)\citenamefont {Paier},
  \citenamefont {Ren}, \citenamefont {Rinke}, \citenamefont {Scuseria},
  \citenamefont {Gr{\"u}neis}, \citenamefont {Kresse},\ and\ \citenamefont
  {Scheffler}}]{paier_assessment_2012}%
  \BibitemOpen
  \bibfield  {author} {\bibinfo {author} {\bibfnamefont {J.}~\bibnamefont
  {Paier}}, \bibinfo {author} {\bibfnamefont {X.}~\bibnamefont {Ren}}, \bibinfo
  {author} {\bibfnamefont {P.}~\bibnamefont {Rinke}}, \bibinfo {author}
  {\bibfnamefont {G.~E.}\ \bibnamefont {Scuseria}}, \bibinfo {author}
  {\bibfnamefont {A.}~\bibnamefont {Gr{\"u}neis}}, \bibinfo {author}
  {\bibfnamefont {G.}~\bibnamefont {Kresse}}, \ and\ \bibinfo {author}
  {\bibfnamefont {M.}~\bibnamefont {Scheffler}},\ }\href {\doibase
  10.1088/1367-2630/14/4/043002} {\bibfield  {journal} {\bibinfo  {journal}
  {New J. Phys.}\ }\textbf {\bibinfo {volume} {14}},\ \bibinfo {pages} {043002}
  (\bibinfo {year} {2012})}\BibitemShut {NoStop}%
\bibitem [{\citenamefont {Bishop}\ and\ \citenamefont
  {L{\"u}hrmann}(1978)}]{bishop_electron_1978}%
  \BibitemOpen
  \bibfield  {author} {\bibinfo {author} {\bibfnamefont {R.~F.}\ \bibnamefont
  {Bishop}}\ and\ \bibinfo {author} {\bibfnamefont {K.~H.}\ \bibnamefont
  {L{\"u}hrmann}},\ }\href {\doibase 10.1103/PhysRevB.17.3757} {\bibfield
  {journal} {\bibinfo  {journal} {Phys. Rev. B}\ }\textbf {\bibinfo {volume}
  {17}},\ \bibinfo {pages} {3757} (\bibinfo {year} {1978})}\BibitemShut
  {NoStop}%
\bibitem [{\citenamefont {Bishop}\ and\ \citenamefont
  {L{\"u}hrmann}(1982)}]{bishop_electron_1982}%
  \BibitemOpen
  \bibfield  {author} {\bibinfo {author} {\bibfnamefont {R.~F.}\ \bibnamefont
  {Bishop}}\ and\ \bibinfo {author} {\bibfnamefont {K.~H.}\ \bibnamefont
  {L{\"u}hrmann}},\ }\href {\doibase 10.1103/PhysRevB.26.5523} {\bibfield
  {journal} {\bibinfo  {journal} {Phys. Rev. B}\ }\textbf {\bibinfo {volume}
  {26}},\ \bibinfo {pages} {5523} (\bibinfo {year} {1982})}\BibitemShut
  {NoStop}%
\bibitem [{\citenamefont {Bishop}(1991)}]{bishop_overview_1991}%
  \BibitemOpen
  \bibfield  {author} {\bibinfo {author} {\bibfnamefont {R.~F.}\ \bibnamefont
  {Bishop}},\ }\href {\doibase 10.1007/BF01119617} {\bibfield  {journal}
  {\bibinfo  {journal} {Theoretica chimica acta}\ }\textbf {\bibinfo {volume}
  {80}},\ \bibinfo {pages} {95} (\bibinfo {year} {1991})}\BibitemShut {NoStop}%
\bibitem [{\citenamefont {Freeman}(1983)}]{freeman_coupled-cluster_1983}%
  \BibitemOpen
  \bibfield  {author} {\bibinfo {author} {\bibfnamefont {D.~L.}\ \bibnamefont
  {Freeman}},\ }\href {\doibase 10.1088/0022-3719/16/4/017} {\bibfield
  {journal} {\bibinfo  {journal} {Journal of Physics C: Solid State Physics}\
  }\textbf {\bibinfo {volume} {16}},\ \bibinfo {pages} {711} (\bibinfo {year}
  {1983})}\BibitemShut {NoStop}%
\bibitem [{\citenamefont {Cioslowski}\ and\ \citenamefont
  {Ziesche}(2005)}]{cioslowski_applicability_2005}%
  \BibitemOpen
  \bibfield  {author} {\bibinfo {author} {\bibfnamefont {J.}~\bibnamefont
  {Cioslowski}}\ and\ \bibinfo {author} {\bibfnamefont {P.}~\bibnamefont
  {Ziesche}},\ }\href {\doibase 10.1103/PhysRevB.71.125105} {\bibfield
  {journal} {\bibinfo  {journal} {Phys. Rev. B}\ }\textbf {\bibinfo {volume}
  {71}},\ \bibinfo {pages} {125105} (\bibinfo {year} {2005})}\BibitemShut
  {NoStop}%
\bibitem [{\citenamefont {Drummond}\ and\ \citenamefont
  {Needs}(2009)}]{drummond_quantum_2009}%
  \BibitemOpen
  \bibfield  {author} {\bibinfo {author} {\bibfnamefont {N.~D.}\ \bibnamefont
  {Drummond}}\ and\ \bibinfo {author} {\bibfnamefont {R.~J.}\ \bibnamefont
  {Needs}},\ }\href {\doibase 10.1103/PhysRevB.79.085414} {\bibfield  {journal}
  {\bibinfo  {journal} {Phys. Rev. B}\ }\textbf {\bibinfo {volume} {79}},\
  \bibinfo {pages} {085414} (\bibinfo {year} {2009})}\BibitemShut {NoStop}%
\bibitem [{\citenamefont {Scuseria}\ \emph {et~al.}(2008)\citenamefont
  {Scuseria}, \citenamefont {Henderson},\ and\ \citenamefont
  {Sorensen}}]{scuseria_ground_2008}%
  \BibitemOpen
  \bibfield  {author} {\bibinfo {author} {\bibfnamefont {G.~E.}\ \bibnamefont
  {Scuseria}}, \bibinfo {author} {\bibfnamefont {T.~M.}\ \bibnamefont
  {Henderson}}, \ and\ \bibinfo {author} {\bibfnamefont {D.~C.}\ \bibnamefont
  {Sorensen}},\ }\href {\doibase doi:10.1063/1.3043729} {\bibfield  {journal}
  {\bibinfo  {journal} {J. Chem. Phys.}\ }\textbf {\bibinfo {volume} {129}},\
  \bibinfo {pages} {231101} (\bibinfo {year} {2008})}\BibitemShut {NoStop}%
\bibitem [{\citenamefont {Peng}\ \emph {et~al.}(2013)\citenamefont {Peng},
  \citenamefont {Steinmann}, \citenamefont {van Aggelen},\ and\ \citenamefont
  {Yang}}]{peng_equivalence_2013}%
  \BibitemOpen
  \bibfield  {author} {\bibinfo {author} {\bibfnamefont {D.}~\bibnamefont
  {Peng}}, \bibinfo {author} {\bibfnamefont {S.~N.}\ \bibnamefont {Steinmann}},
  \bibinfo {author} {\bibfnamefont {H.}~\bibnamefont {van Aggelen}}, \ and\
  \bibinfo {author} {\bibfnamefont {W.}~\bibnamefont {Yang}},\ }\href {\doibase
  doi:10.1063/1.4820556} {\bibfield  {journal} {\bibinfo  {journal} {J. Chem.
  Phys.}\ }\textbf {\bibinfo {volume} {139}},\ \bibinfo {pages} {104112}
  (\bibinfo {year} {2013})}\BibitemShut {NoStop}%
\bibitem [{\citenamefont {Scuseria}\ \emph {et~al.}(2013)\citenamefont
  {Scuseria}, \citenamefont {Henderson},\ and\ \citenamefont
  {Bulik}}]{scuseria_particle-particle_2013}%
  \BibitemOpen
  \bibfield  {author} {\bibinfo {author} {\bibfnamefont {G.~E.}\ \bibnamefont
  {Scuseria}}, \bibinfo {author} {\bibfnamefont {T.~M.}\ \bibnamefont
  {Henderson}}, \ and\ \bibinfo {author} {\bibfnamefont {I.~W.}\ \bibnamefont
  {Bulik}},\ }\href {\doibase doi:10.1063/1.4820557} {\bibfield  {journal}
  {\bibinfo  {journal} {J. Chem. Phys.}\ }\textbf {\bibinfo {volume} {139}},\
  \bibinfo {pages} {104113} (\bibinfo {year} {2013})}\BibitemShut {NoStop}%
\bibitem [{\citenamefont {van Aggelen}\ \emph {et~al.}(2013)\citenamefont {van
  Aggelen}, \citenamefont {Yang},\ and\ \citenamefont
  {Yang}}]{van_aggelen_exchange-correlation_2013}%
  \BibitemOpen
  \bibfield  {author} {\bibinfo {author} {\bibfnamefont {H.}~\bibnamefont {van
  Aggelen}}, \bibinfo {author} {\bibfnamefont {Y.}~\bibnamefont {Yang}}, \ and\
  \bibinfo {author} {\bibfnamefont {W.}~\bibnamefont {Yang}},\ }\href {\doibase
  10.1103/PhysRevA.88.030501} {\bibfield  {journal} {\bibinfo  {journal} {Phys.
  Rev. A}\ }\textbf {\bibinfo {volume} {88}},\ \bibinfo {pages} {030501}
  (\bibinfo {year} {2013})}\BibitemShut {NoStop}%
\bibitem [{\citenamefont {Ceperley}\ and\ \citenamefont
  {Alder}(1980)}]{ceperley_ground_1980}%
  \BibitemOpen
  \bibfield  {author} {\bibinfo {author} {\bibfnamefont {D.~M.}\ \bibnamefont
  {Ceperley}}\ and\ \bibinfo {author} {\bibfnamefont {B.~J.}\ \bibnamefont
  {Alder}},\ }\href {\doibase 10.1103/PhysRevLett.45.566} {\bibfield  {journal}
  {\bibinfo  {journal} {Phys. Rev. Lett.}\ }\textbf {\bibinfo {volume} {45}},\
  \bibinfo {pages} {566} (\bibinfo {year} {1980})}\BibitemShut {NoStop}%
\bibitem [{\citenamefont {Perdew}\ and\ \citenamefont
  {Zunger}(1981)}]{perdew_self-interaction_1981}%
  \BibitemOpen
  \bibfield  {author} {\bibinfo {author} {\bibfnamefont {J.~P.}\ \bibnamefont
  {Perdew}}\ and\ \bibinfo {author} {\bibfnamefont {A.}~\bibnamefont
  {Zunger}},\ }\href {\doibase 10.1103/PhysRevB.23.5048} {\bibfield  {journal}
  {\bibinfo  {journal} {Phys. Rev. B}\ }\textbf {\bibinfo {volume} {23}},\
  \bibinfo {pages} {5048} (\bibinfo {year} {1981})}\BibitemShut {NoStop}%
\bibitem [{\citenamefont {Huotari}\ \emph {et~al.}(2010)\citenamefont
  {Huotari}, \citenamefont {Soininen}, \citenamefont {Pylkk{\"a}nen},
  \citenamefont {H{\"a}m{\"a}l{\"a}inen}, \citenamefont {Issolah},
  \citenamefont {Titov}, \citenamefont {{McMinis}}, \citenamefont {Kim},
  \citenamefont {Esler}, \citenamefont {Ceperley}, \citenamefont {Holzmann},\
  and\ \citenamefont {Olevano}}]{huotari_momentum_2010}%
  \BibitemOpen
  \bibfield  {author} {\bibinfo {author} {\bibfnamefont {S.}~\bibnamefont
  {Huotari}}, \bibinfo {author} {\bibfnamefont {J.~A.}\ \bibnamefont
  {Soininen}}, \bibinfo {author} {\bibfnamefont {T.}~\bibnamefont
  {Pylkk{\"a}nen}}, \bibinfo {author} {\bibfnamefont {K.}~\bibnamefont
  {H{\"a}m{\"a}l{\"a}inen}}, \bibinfo {author} {\bibfnamefont {A.}~\bibnamefont
  {Issolah}}, \bibinfo {author} {\bibfnamefont {A.}~\bibnamefont {Titov}},
  \bibinfo {author} {\bibfnamefont {J.}~\bibnamefont {{McMinis}}}, \bibinfo
  {author} {\bibfnamefont {J.}~\bibnamefont {Kim}}, \bibinfo {author}
  {\bibfnamefont {K.}~\bibnamefont {Esler}}, \bibinfo {author} {\bibfnamefont
  {D.~M.}\ \bibnamefont {Ceperley}}, \bibinfo {author} {\bibfnamefont
  {M.}~\bibnamefont {Holzmann}}, \ and\ \bibinfo {author} {\bibfnamefont
  {V.}~\bibnamefont {Olevano}},\ }\href {\doibase
  10.1103/PhysRevLett.105.086403} {\bibfield  {journal} {\bibinfo  {journal}
  {Phys. Rev. Lett.}\ }\textbf {\bibinfo {volume} {105}},\ \bibinfo {pages}
  {086403} (\bibinfo {year} {2010})}\BibitemShut {NoStop}%
\bibitem [{\citenamefont {Drummond}\ \emph {et~al.}(2011)\citenamefont
  {Drummond}, \citenamefont {L{\'o}pez~R{\'i}os}, \citenamefont {Needs},\ and\
  \citenamefont {Pickard}}]{drummond_quantum_2011}%
  \BibitemOpen
  \bibfield  {author} {\bibinfo {author} {\bibfnamefont {N.~D.}\ \bibnamefont
  {Drummond}}, \bibinfo {author} {\bibfnamefont {P.}~\bibnamefont
  {L{\'o}pez~R{\'i}os}}, \bibinfo {author} {\bibfnamefont {R.~J.}\ \bibnamefont
  {Needs}}, \ and\ \bibinfo {author} {\bibfnamefont {C.~J.}\ \bibnamefont
  {Pickard}},\ }\href {\doibase 10.1103/PhysRevLett.107.207402} {\bibfield
  {journal} {\bibinfo  {journal} {Phys. Rev. Lett.}\ }\textbf {\bibinfo
  {volume} {107}},\ \bibinfo {pages} {207402} (\bibinfo {year}
  {2011})}\BibitemShut {NoStop}%
\bibitem [{\citenamefont {Holzmann}\ \emph {et~al.}(2011)\citenamefont
  {Holzmann}, \citenamefont {Bernu}, \citenamefont {Pierleoni}, \citenamefont
  {{McMinis}}, \citenamefont {Ceperley}, \citenamefont {Olevano},\ and\
  \citenamefont {Delle~Site}}]{holzmann_momentum_2011}%
  \BibitemOpen
  \bibfield  {author} {\bibinfo {author} {\bibfnamefont {M.}~\bibnamefont
  {Holzmann}}, \bibinfo {author} {\bibfnamefont {B.}~\bibnamefont {Bernu}},
  \bibinfo {author} {\bibfnamefont {C.}~\bibnamefont {Pierleoni}}, \bibinfo
  {author} {\bibfnamefont {J.}~\bibnamefont {{McMinis}}}, \bibinfo {author}
  {\bibfnamefont {D.~M.}\ \bibnamefont {Ceperley}}, \bibinfo {author}
  {\bibfnamefont {V.}~\bibnamefont {Olevano}}, \ and\ \bibinfo {author}
  {\bibfnamefont {L.}~\bibnamefont {Delle~Site}},\ }\href {\doibase
  10.1103/PhysRevLett.107.110402} {\bibfield  {journal} {\bibinfo  {journal}
  {Phys. Rev. Lett.}\ }\textbf {\bibinfo {volume} {107}},\ \bibinfo {pages}
  {110402} (\bibinfo {year} {2011})}\BibitemShut {NoStop}%
\bibitem [{\citenamefont {Baguet}\ \emph {et~al.}(2013)\citenamefont {Baguet},
  \citenamefont {Delyon}, \citenamefont {Bernu},\ and\ \citenamefont
  {Holzmann}}]{baguet_hartree-fock_2013}%
  \BibitemOpen
  \bibfield  {author} {\bibinfo {author} {\bibfnamefont {L.}~\bibnamefont
  {Baguet}}, \bibinfo {author} {\bibfnamefont {F.}~\bibnamefont {Delyon}},
  \bibinfo {author} {\bibfnamefont {B.}~\bibnamefont {Bernu}}, \ and\ \bibinfo
  {author} {\bibfnamefont {M.}~\bibnamefont {Holzmann}},\ }\href {\doibase
  10.1103/PhysRevLett.111.166402} {\bibfield  {journal} {\bibinfo  {journal}
  {Phys. Rev. Lett.}\ }\textbf {\bibinfo {volume} {111}},\ \bibinfo {pages}
  {166402} (\bibinfo {year} {2013})}\BibitemShut {NoStop}%
\bibitem [{\citenamefont {Bartlett}\ and\ \citenamefont {Musia{\l
  }}(2007)}]{bartlett_coupled-cluster_2007}%
  \BibitemOpen
  \bibfield  {author} {\bibinfo {author} {\bibfnamefont {R.~J.}\ \bibnamefont
  {Bartlett}}\ and\ \bibinfo {author} {\bibfnamefont {M.}~\bibnamefont
  {Musia{\l }}},\ }\href {\doibase 10.1103/RevModPhys.79.291} {\bibfield
  {journal} {\bibinfo  {journal} {Rev. Mod. Phys.}\ }\textbf {\bibinfo {volume}
  {79}},\ \bibinfo {pages} {291} (\bibinfo {year} {2007})}\BibitemShut
  {NoStop}%
\bibitem [{\citenamefont {Ring}\ and\ \citenamefont
  {Schuck}(1980)}]{ring_nuclear_1980}%
  \BibitemOpen
  \bibfield  {author} {\bibinfo {author} {\bibfnamefont {P.}~\bibnamefont
  {Ring}}\ and\ \bibinfo {author} {\bibfnamefont {P.}~\bibnamefont {Schuck}},\
  }\href@noop {} {\emph {\bibinfo {title} {The nuclear many-body problem}}}\
  (\bibinfo  {publisher} {Springer-Verlag},\ \bibinfo {address} {Berlin; New
  York},\ \bibinfo {year} {1980})\BibitemShut {NoStop}%
\bibitem [{\citenamefont {Scuseria}(1995)}]{scuseria_connections_1995}%
  \BibitemOpen
  \bibfield  {author} {\bibinfo {author} {\bibfnamefont {G.~E.}\ \bibnamefont
  {Scuseria}},\ }\href {\doibase 10.1002/qua.560550211} {\bibfield  {journal}
  {\bibinfo  {journal} {Int. J. Quant. Chem.}\ }\textbf {\bibinfo {volume}
  {55}},\ \bibinfo {pages} {165{\textendash}171} (\bibinfo {year}
  {1995})}\BibitemShut {NoStop}%
\bibitem [{\citenamefont {Shimazaki}\ and\ \citenamefont
  {Asai}(2008)}]{shimazaki_band_2008}%
  \BibitemOpen
  \bibfield  {author} {\bibinfo {author} {\bibfnamefont {T.}~\bibnamefont
  {Shimazaki}}\ and\ \bibinfo {author} {\bibfnamefont {Y.}~\bibnamefont
  {Asai}},\ }\href {\doibase 10.1016/j.cplett.2008.10.012} {\bibfield
  {journal} {\bibinfo  {journal} {Chem. Phys. Lett.}\ }\textbf {\bibinfo
  {volume} {466}},\ \bibinfo {pages} {91} (\bibinfo {year} {2008})}\BibitemShut
  {NoStop}%
\bibitem [{Note1()}]{Note1}%
  \BibitemOpen
  \bibinfo {note} {See supplementary information.}\BibitemShut {Stop}%
\bibitem [{\citenamefont {Drummond}\ \emph {et~al.}(2008)\citenamefont
  {Drummond}, \citenamefont {Needs}, \citenamefont {Sorouri},\ and\
  \citenamefont {Foulkes}}]{drummond_finite-size_2008}%
  \BibitemOpen
  \bibfield  {author} {\bibinfo {author} {\bibfnamefont {N.~D.}\ \bibnamefont
  {Drummond}}, \bibinfo {author} {\bibfnamefont {R.~J.}\ \bibnamefont {Needs}},
  \bibinfo {author} {\bibfnamefont {A.}~\bibnamefont {Sorouri}}, \ and\
  \bibinfo {author} {\bibfnamefont {W.~M.~C.}\ \bibnamefont {Foulkes}},\ }\href
  {\doibase 10.1103/PhysRevB.78.125106} {\bibfield  {journal} {\bibinfo
  {journal} {Phys. Rev. B}\ }\textbf {\bibinfo {volume} {78}},\ \bibinfo
  {pages} {125106} (\bibinfo {year} {2008})}\BibitemShut {NoStop}%
\bibitem [{\citenamefont {Shepherd}\ \emph
  {et~al.}(2012{\natexlab{a}})\citenamefont {Shepherd}, \citenamefont
  {Gr{\"u}neis}, \citenamefont {Booth}, \citenamefont {Kresse},\ and\
  \citenamefont {Alavi}}]{shepherd_convergence_2012}%
  \BibitemOpen
  \bibfield  {author} {\bibinfo {author} {\bibfnamefont {J.~J.}\ \bibnamefont
  {Shepherd}}, \bibinfo {author} {\bibfnamefont {A.}~\bibnamefont
  {Gr{\"u}neis}}, \bibinfo {author} {\bibfnamefont {G.~H.}\ \bibnamefont
  {Booth}}, \bibinfo {author} {\bibfnamefont {G.}~\bibnamefont {Kresse}}, \
  and\ \bibinfo {author} {\bibfnamefont {A.}~\bibnamefont {Alavi}},\ }\href
  {\doibase 10.1103/PhysRevB.86.035111} {\bibfield  {journal} {\bibinfo
  {journal} {Phys. Rev. B}\ }\textbf {\bibinfo {volume} {86}},\ \bibinfo
  {pages} {035111} (\bibinfo {year} {2012}{\natexlab{a}})}\BibitemShut
  {NoStop}%
\bibitem [{\citenamefont {Gr{\"u}neis}\ \emph {et~al.}(2013)\citenamefont
  {Gr{\"u}neis}, \citenamefont {Shepherd}, \citenamefont {Alavi}, \citenamefont
  {Tew},\ and\ \citenamefont {Booth}}]{gruneis_explicitly_2013}%
  \BibitemOpen
  \bibfield  {author} {\bibinfo {author} {\bibfnamefont {A.}~\bibnamefont
  {Gr{\"u}neis}}, \bibinfo {author} {\bibfnamefont {J.~J.}\ \bibnamefont
  {Shepherd}}, \bibinfo {author} {\bibfnamefont {A.}~\bibnamefont {Alavi}},
  \bibinfo {author} {\bibfnamefont {D.~P.}\ \bibnamefont {Tew}}, \ and\
  \bibinfo {author} {\bibfnamefont {G.~H.}\ \bibnamefont {Booth}},\ }\href
  {\doibase doi:10.1063/1.4818753} {\bibfield  {journal} {\bibinfo  {journal}
  {J. Chem. Phys.}\ }\textbf {\bibinfo {volume} {139}},\ \bibinfo {pages}
  {084112} (\bibinfo {year} {2013})}\BibitemShut {NoStop}%
\bibitem [{\citenamefont {H{\"a}ttig}\ \emph {et~al.}(2012)\citenamefont
  {H{\"a}ttig}, \citenamefont {Klopper}, \citenamefont {K{\"o}hn},\ and\
  \citenamefont {Tew}}]{hattig_explicitly_2012}%
  \BibitemOpen
  \bibfield  {author} {\bibinfo {author} {\bibfnamefont {C.}~\bibnamefont
  {H{\"a}ttig}}, \bibinfo {author} {\bibfnamefont {W.}~\bibnamefont {Klopper}},
  \bibinfo {author} {\bibfnamefont {A.}~\bibnamefont {K{\"o}hn}}, \ and\
  \bibinfo {author} {\bibfnamefont {D.~P.}\ \bibnamefont {Tew}},\ }\href
  {\doibase 10.1021/cr200168z} {\bibfield  {journal} {\bibinfo  {journal}
  {Chem. Rev.}\ }\textbf {\bibinfo {volume} {112}},\ \bibinfo {pages} {4}
  (\bibinfo {year} {2012})}\BibitemShut {NoStop}%
\bibitem [{Note2()}]{Note2}%
  \BibitemOpen
  \bibinfo {note} {For $N=14$, the benchmarks here come from full configuration
  interaction quantum Monte Carlo~\cite {booth_towards_2013} for the
  high-to-metallic density regime, and diffusion Monte Carlo for the
  remainder~\cite {needs_continuum_2010,lopez_rios__2013}. These results are
  effectively exact following extensive development and benchmarking~\cite
  {shepherd_full_2012,shepherd_investigation_2012,shepherd_convergence_2012,shepherd_many-body_2013}}\BibitemShut
  {NoStop}%
\bibitem [{\citenamefont {Kwon}\ \emph {et~al.}(1998)\citenamefont {Kwon},
  \citenamefont {Ceperley},\ and\ \citenamefont {Martin}}]{kwon_effects_1998}%
  \BibitemOpen
  \bibfield  {author} {\bibinfo {author} {\bibfnamefont {Y.}~\bibnamefont
  {Kwon}}, \bibinfo {author} {\bibfnamefont {D.~M.}\ \bibnamefont {Ceperley}},
  \ and\ \bibinfo {author} {\bibfnamefont {R.~M.}\ \bibnamefont {Martin}},\
  }\href {\doibase 10.1103/PhysRevB.58.6800} {\bibfield  {journal} {\bibinfo
  {journal} {Phys. Rev. B}\ }\textbf {\bibinfo {volume} {58}},\ \bibinfo
  {pages} {6800} (\bibinfo {year} {1998})}\BibitemShut {NoStop}%
\bibitem [{\citenamefont {L{\'o}pez~R{\'i}os}\ \emph
  {et~al.}(2006)\citenamefont {L{\'o}pez~R{\'i}os}, \citenamefont {Ma},
  \citenamefont {Drummond}, \citenamefont {Towler},\ and\ \citenamefont
  {Needs}}]{lopez_rios_inhomogeneous_2006}%
  \BibitemOpen
  \bibfield  {author} {\bibinfo {author} {\bibfnamefont {P.}~\bibnamefont
  {L{\'o}pez~R{\'i}os}}, \bibinfo {author} {\bibfnamefont {A.}~\bibnamefont
  {Ma}}, \bibinfo {author} {\bibfnamefont {N.~D.}\ \bibnamefont {Drummond}},
  \bibinfo {author} {\bibfnamefont {M.~D.}\ \bibnamefont {Towler}}, \ and\
  \bibinfo {author} {\bibfnamefont {R.~J.}\ \bibnamefont {Needs}},\ }\href
  {\doibase 10.1103/PhysRevE.74.066701} {\bibfield  {journal} {\bibinfo
  {journal} {Phys. Rev. E}\ }\textbf {\bibinfo {volume} {74}},\ \bibinfo
  {pages} {066701} (\bibinfo {year} {2006})}\BibitemShut {NoStop}%
\bibitem [{\citenamefont {Shepherd}\ \emph
  {et~al.}(2012{\natexlab{b}})\citenamefont {Shepherd}, \citenamefont {Booth},
  \citenamefont {Gr{\"u}neis},\ and\ \citenamefont
  {Alavi}}]{shepherd_full_2012}%
  \BibitemOpen
  \bibfield  {author} {\bibinfo {author} {\bibfnamefont {J.~J.}\ \bibnamefont
  {Shepherd}}, \bibinfo {author} {\bibfnamefont {G.}~\bibnamefont {Booth}},
  \bibinfo {author} {\bibfnamefont {A.}~\bibnamefont {Gr{\"u}neis}}, \ and\
  \bibinfo {author} {\bibfnamefont {A.}~\bibnamefont {Alavi}},\ }\href
  {\doibase 10.1103/PhysRevB.85.081103} {\bibfield  {journal} {\bibinfo
  {journal} {Phys. Rev. B}\ }\textbf {\bibinfo {volume} {85}},\ \bibinfo
  {pages} {081103} (\bibinfo {year} {2012}{\natexlab{b}})}\BibitemShut
  {NoStop}%
\bibitem [{\citenamefont {Shepherd}\ \emph
  {et~al.}(2012{\natexlab{c}})\citenamefont {Shepherd}, \citenamefont {Booth},\
  and\ \citenamefont {Alavi}}]{shepherd_investigation_2012}%
  \BibitemOpen
  \bibfield  {author} {\bibinfo {author} {\bibfnamefont {J.~J.}\ \bibnamefont
  {Shepherd}}, \bibinfo {author} {\bibfnamefont {G.~H.}\ \bibnamefont {Booth}},
  \ and\ \bibinfo {author} {\bibfnamefont {A.}~\bibnamefont {Alavi}},\ }\href
  {\doibase 10.1063/1.4720076} {\bibfield  {journal} {\bibinfo  {journal} {J.
  Chem. Phys.}\ }\textbf {\bibinfo {volume} {136}},\ \bibinfo {pages} {244101}
  (\bibinfo {year} {2012}{\natexlab{c}})}\BibitemShut {NoStop}%
\bibitem [{\citenamefont {Gr{\"u}neis}\ \emph {et~al.}(2009)\citenamefont
  {Gr{\"u}neis}, \citenamefont {Marsman}, \citenamefont {Harl}, \citenamefont
  {Schimka},\ and\ \citenamefont {Kresse}}]{gruneis_making_2009}%
  \BibitemOpen
  \bibfield  {author} {\bibinfo {author} {\bibfnamefont {A.}~\bibnamefont
  {Gr{\"u}neis}}, \bibinfo {author} {\bibfnamefont {M.}~\bibnamefont
  {Marsman}}, \bibinfo {author} {\bibfnamefont {J.}~\bibnamefont {Harl}},
  \bibinfo {author} {\bibfnamefont {L.}~\bibnamefont {Schimka}}, \ and\
  \bibinfo {author} {\bibfnamefont {G.}~\bibnamefont {Kresse}},\ }\href
  {\doibase 10.1063/1.3250347} {\bibfield  {journal} {\bibinfo  {journal} {J.
  Chem. Phys.}\ }\textbf {\bibinfo {volume} {131}},\ \bibinfo {pages} {154115}
  (\bibinfo {year} {2009})}\BibitemShut {NoStop}%
\bibitem [{\citenamefont {Shepherd}\ and\ \citenamefont
  {Gr{\"u}neis}(2013)}]{shepherd_many-body_2013}%
  \BibitemOpen
  \bibfield  {author} {\bibinfo {author} {\bibfnamefont {J.~J.}\ \bibnamefont
  {Shepherd}}\ and\ \bibinfo {author} {\bibfnamefont {A.}~\bibnamefont
  {Gr{\"u}neis}},\ }\href {\doibase 10.1103/PhysRevLett.110.226401} {\bibfield
  {journal} {\bibinfo  {journal} {Phys. Rev. Lett.}\ }\textbf {\bibinfo
  {volume} {110}},\ \bibinfo {pages} {226401} (\bibinfo {year}
  {2013})}\BibitemShut {NoStop}%
\bibitem [{\citenamefont {Shepherd}\ \emph
  {et~al.}(2013{\natexlab{a}})\citenamefont {Shepherd}, \citenamefont
  {Henderson},\ and\ \citenamefont {Scuseria}}]{shepherd_coupled_2013}%
  \BibitemOpen
  \bibfield  {author} {\bibinfo {author} {\bibfnamefont {J.~J.}\ \bibnamefont
  {Shepherd}}, \bibinfo {author} {\bibfnamefont {T.~M.}\ \bibnamefont
  {Henderson}}, \ and\ \bibinfo {author} {\bibfnamefont {G.~E.}\ \bibnamefont
  {Scuseria}},\ }\href {http://arxiv.org/abs/1310.6806} {\bibfield  {journal}
  {\bibinfo  {journal} {{arXiv:1310.6806}}\ } (\bibinfo {year}
  {2013}{\natexlab{a}})}\BibitemShut {NoStop}%
\bibitem [{\citenamefont {Shepherd}\ \emph
  {et~al.}(2013{\natexlab{b}})\citenamefont {Shepherd}, \citenamefont
  {R{\'i}os}, \citenamefont {Drummond}, \citenamefont {Needs},\ and\
  \citenamefont {Alavi}}]{shepherd__2013}%
  \BibitemOpen
  \bibfield  {author} {\bibinfo {author} {\bibfnamefont {J.~J.}\ \bibnamefont
  {Shepherd}}, \bibinfo {author} {\bibfnamefont {P.~L.}\ \bibnamefont
  {R{\'i}os}}, \bibinfo {author} {\bibfnamefont {N.~D.}\ \bibnamefont
  {Drummond}}, \bibinfo {author} {\bibfnamefont {R.~J.}\ \bibnamefont {Needs}},
  \ and\ \bibinfo {author} {\bibfnamefont {A.}~\bibnamefont {Alavi}},\
  }\href@noop {} {}\bibinfo {type} {In preparation}\ (\bibinfo {year}
  {2013})\BibitemShut {NoStop}%
\bibitem [{\citenamefont {Booth}\ \emph {et~al.}(2013)\citenamefont {Booth},
  \citenamefont {Gr{\"u}neis}, \citenamefont {Kresse},\ and\ \citenamefont
  {Alavi}}]{booth_towards_2013}%
  \BibitemOpen
  \bibfield  {author} {\bibinfo {author} {\bibfnamefont {G.~H.}\ \bibnamefont
  {Booth}}, \bibinfo {author} {\bibfnamefont {A.}~\bibnamefont {Gr{\"u}neis}},
  \bibinfo {author} {\bibfnamefont {G.}~\bibnamefont {Kresse}}, \ and\ \bibinfo
  {author} {\bibfnamefont {A.}~\bibnamefont {Alavi}},\ }\href {\doibase
  10.1038/nature11770} {\bibfield  {journal} {\bibinfo  {journal} {Nature}\
  }\textbf {\bibinfo {volume} {493}},\ \bibinfo {pages} {365} (\bibinfo {year}
  {2013})}\BibitemShut {NoStop}%
\bibitem [{\citenamefont {Needs}\ \emph {et~al.}(2010)\citenamefont {Needs},
  \citenamefont {Towler}, \citenamefont {Drummond},\ and\ \citenamefont
  {L{\'o}pez~R{\'i}os}}]{needs_continuum_2010}%
  \BibitemOpen
  \bibfield  {author} {\bibinfo {author} {\bibfnamefont {R.~J.}\ \bibnamefont
  {Needs}}, \bibinfo {author} {\bibfnamefont {M.~D.}\ \bibnamefont {Towler}},
  \bibinfo {author} {\bibfnamefont {N.~D.}\ \bibnamefont {Drummond}}, \ and\
  \bibinfo {author} {\bibfnamefont {P.}~\bibnamefont {L{\'o}pez~R{\'i}os}},\
  }\href {\doibase 10.1088/0953-8984/22/2/023201} {\bibfield  {journal}
  {\bibinfo  {journal} {J. Phys.: Condens. Matter}\ }\textbf {\bibinfo {volume}
  {22}},\ \bibinfo {pages} {023201} (\bibinfo {year} {2010})}\BibitemShut
  {NoStop}%
\bibitem [{\citenamefont {L{\'o}pez~R{\'i}os}(2013)}]{lopez_rios__2013}%
  \BibitemOpen
  \bibfield  {author} {\bibinfo {author} {\bibfnamefont {P.}~\bibnamefont
  {L{\'o}pez~R{\'i}os}},\ }\href@noop {} {}\bibinfo {type} {Pers. Comm.}\
  (\bibinfo {year} {2013})\BibitemShut {NoStop}%
\end{thebibliography}
\end{document}